\documentclass[12pt,twoside]{article}
\usepackage{amsmath,feynmf}

\newif\iflowmem\lowmemfalse

\DeclareMathOperator*{\tsum}{{\textstyle \sum}}
\DeclareMathOperator*{\tprod}{{\textstyle \prod}}

\def\kslash{k \kern-0.36em \raisebox{0em}{\scriptsize /} 
              \kern 0em}
\def\lslash{\ell \kern-0.36em \raisebox{0em}{\scriptsize /} 
              \kern 0em}
\def\pslash{p \kern-0.34em \raisebox{-0.05em}{\scriptsize /} 
              \kern 0em}
\def\id{\mathrm{id}}

\newtheorem{prp}{Proposition}
\def\qed{\leavevmode \hfill\hbox to.77778em{\hfil\vrule
         \vbox to.675em{\hrule width.6em\vfil\hrule}\vrule\hfil}}
 
\makeatletter

\def\section{\@startsection{section}{1}{\z@}{-3.25ex plus -1ex minus
            -.2ex}{1.5ex plus .2ex}{\normalfont\bfseries}}

\def\thebibliography#1{\section*{\refname}\list
  {\@biblabel{\theenumiv}}{\settowidth\labelwidth{\@biblabel{#1}}%
    \leftmargin\labelwidth
    \advance\leftmargin\labelsep
    \footnotesize \parsep=0pt \itemsep=0pt
    \usecounter{enumiv}%
    \let\p@enumiv\@empty
    \def\theenumiv{\arabic{enumiv}}}%
    \def\newblock{\hskip .11em plus.33em minus.07em}%
    \sloppy\clubpenalty4000\widowpenalty4000
    \sfcode`\.=1000\relax}

\newlength\testa

\def\VL{\setlength{\testa}{1mm}\addtolength{\testa}{\arrayrulewidth}
             \hskip -2pt
             \vrule \@height \testa \@width 1mm \@depth -1mm 
             \vrule \@width  \arrayrulewidth \@height \testa \hskip 2pt}
\def\AL{\setlength{\testa}{1mm}\addtolength{\testa}{\arrayrulewidth}
             \hskip -2pt
             \vrule \@height \testa \@width 1mm \@depth -1mm
             \vrule \@width \arrayrulewidth \@depth -1mm \hskip 2pt}
\def\TL{\setlength{\testa}{1mm}\addtolength{\testa}{\arrayrulewidth}
             \hskip -2pt
             \vrule \@height \testa \@width 1mm \@depth -1mm
             \vrule \@width \arrayrulewidth \hskip 2pt}
\def\IL{\setlength{\testa}{1mm}\addtolength{\testa}{\arrayrulewidth}
             \hskip -2pt 
             \vrule \@height 0pt \@width 1mm \@depth 0pt 
             \vrule \@width \arrayrulewidth \hskip 2pt }

\def\ps@hep{\addtolength{\headheight}{5pt}
            \addtolength{\topmargin}{-15pt}
            \addtolength{\headsep}{15pt}
    \def\@oddhead{\hfil\begin{tabular}{r}
          \texttt{hep-th/9805098} v4\\ CPT-98/P.3639
          \end{tabular}}
          \let\@evenhead\@oddhead
          \def\@oddfoot{\hfil\thepage\hfil}\let\@evenfoot\@oddfoot}

\makeatother

\allowdisplaybreaks[2]
\begin{document}

\thispagestyle{hep}

\vskip 3mm
\begin{center}
{\renewcommand{\thefootnote}{\fnsymbol{footnote}}
{\large\uppercase{On Kreimer's Hopf algebra structure \\[0.5ex]
of Feynman graphs}}
\\[5ex]
{\bfseries Thomas KRAJEWSKI\footnote{and Universit\'e de Provence and
    Ecole Normale Superieure de Lyon, \\ \hspace*{1.4em} 
    \texttt{tkrajews@cpt.univ-mrs.fr}} 
{\normalfont and}
Raimar WULKENHAAR\footnote{and Universit\"at Leipzig, 
  \texttt{raimar@cpt.univ-mrs.fr}}\footnote{supported by the German 
  Academic Exchange Service (DAAD), grant no.\ D/97/20386.}} 
\setcounter{footnote}{0}}
\vskip 4mm

{\itshape Centre de Physique Th\'eorique \\
          CNRS - Luminy, Case 907 \\ 
          13288 Marseille Cedex 9, France} 

\vskip 4mm {\small November 4, 1998}

\end{center}
\vskip 4ex
 
\begin{abstract}
We reinvestigate Kreimer's Hopf algebra structure of perturbative
quantum field theories with a special emphasis on overlapping
divergences. Kreimer first disentangles overlapping divergences into a
linear combination of disjoint and nested ones and then tackles that
linear combination by the Hopf algebra operations. We present a
formulation where the Hopf algebra operations are directly defined on
any type of divergence. We explain the precise relation to Kreimer's
Hopf algebra and obtain thereby a characterization of their primitive
elements.
\end{abstract}

\noindent
{\tabcolsep 0pt
\begin{tabular}{ll}PACS-98:~{} & 
02.10.Sp Linear and multilinear algebra, 
11.10.Gh Renormalization,  \\
&11.15.Bt General properties of perturbation theory \end{tabular}}
\vskip 1ex

\section{Introduction}

This paper is the result of our efforts to understand the article by
Dirk Kreimer on the Hopf algebra structure of perturbative quantum
field theories \cite{k}. That article was brought to our attention by
Alain Connes in his talk during the \mbox{Vietri} conference on
noncommutative geometry. Kreimer discovered that divergent Feynman
graphs can be understood as elements of a Hopf algebra. The forest
formula guiding the renormalization of Feynman graphs with
subdivergences is reproduced by a certain interplay of product,
coproduct, antipode and counit of that Hopf algebra. Meanwhile Connes
and Kreimer elaborated a deep structural link \cite{ck} between that
Hopf algebra of renormalization and the Hopf algebra emerging in the
computation of the local index formula for transverse hypoelliptic
operators \cite{cm}. This indicates that renormalization provides a
mathematical calculus that can be thought of as a refinement of
diffeomorphisms. 

As explained by Kreimer in \cite{k,ck,k3} and in private discussions, 
overlapping divergences require a special treatment. 
Overlapping divergences must first be disentangled into a linear
combination of terms containing disjoint or nested divergences
exclusively. Suppose the (divergent) integrand $I(q_1,\dots, q_n)$
corresponding to a Feynman graph depends on $n$ external
parameters (masses and momenta) $q_i,\dots,q_n$. The idea is to write 
\[
I(q_1,q_2,\dots, q_n)
=\{I(q_1,q_2,\dots, q_n)-I(q_1,0,\dots, 0)\} + I(q_1,0,\dots, 0)~.
\]
The integrand $\{I(q_1,q_2,\dots, q_n) -I(q_1,0,\dots, 0)\}$ is less
divergent, in the optimal case convergent or without
subdivergences. It is therefore sufficient to consider integrands
depending on a single scale $q$. In the same way as above one can
write $I(q)=\{I(q)-I'(q)\}+I(q)$, where $I'(q)$ is derived from $I(q)$
by nullifying $q$ in some parts of $I(q)$. It was shown in \cite{kh}
that by this procedure (which is encoded in the Schwinger-Dyson
equation) it is always possible to disentangle overlapping
divergences. Hence one can restrict the operations of the Hopf algebra
to terms containing no overlapping divergences. The forest
formula is trivial in this case, it simply says that the
subdivergences must be compensated in ascending order.

In this paper we present our independent approach to the problem of
overlapping divergences. Our goal is to treat overlapping divergences
on the same footing with disjoint and nested ones so that the
operations of the Hopf algebra are directly defined on any Feynman
graph. We show that this aim can be achieved by endowing Kreimer's
parenthesized words (PW) describing the Feynman graphs with additional
information. In our formulation, a PW is a collection of all maximal
forests of a Feynman graph, where identical regions in various forests
are visualized. We show that one of the antipode axioms recovers the
forest formula in its full beauty for any Feynman graph. Following an
idea by Dirk Kreimer \cite{ck,k3} we describe the precise relation
between his and our formulations of the Hopf algebra of
renormalization. In this way we gain an explicit construction of those
primitive elements of Kreimer's Hopf algebra which are different from
the graphically primitive elements.

Our paper is organized as follows: We introduce in section \ref{for}
our extended PWs and discuss in section \ref{ro} the $R$-operation of
renormalization. The Hopf algebra is identified in section \ref{hopf},
where longer proofs are delegated to the appendix. In section
\ref{KWK} we discuss the relation to the Hopf algebra of Kreimer. In
sections \ref{ex1} and \ref{ex2} we apply our methods to examples of
Feynman graphs with overlapping divergences.

\section{Feynman graphs, maximal forests and parenthesized words}
\label{for}

\iflowmem
  \begin{fmffile}{fmfhopf1}\else
  \begin{fmffile}{fmfhopf}
\fi

Let $\Gamma$ be a Feynman graph. In the way described by Kreimer we
draw boxes around superficially (UV-) divergent sectors of $\Gamma$: 
\begin{equation}
\label{bsp}
\parbox{60mm}{\begin{picture}(60,35)
\put(0,0){\begin{fmfgraph}(60,35)
\fmfleft{l}
\fmfright{r1,r2}
  \fmf{photon,tension=3}{l,i}
  \fmf{fermion,tension=1.5}{r1,b2}
  \fmf{fermion,tension=0.8}{b2,b1}
  \fmf{fermion}{b1,i}
  \fmf{fermion}{i,t1}
  \fmf{fermion,tension=0.8}{t1,t2}
  \fmf{fermion,tension=1.5}{t2,r2}
\fmffreeze
  \fmf{photon}{t1,b1}
  \fmf{photon}{t2,p1}
   \fmf{phantom,tension=1.5}{p1,p2}
  \fmf{photon}{p2,b2}
\fmffreeze
  \fmf{fermion,left}{p1,p2}
  \fmf{fermion,left}{p2,p1}
     \fmf{phantom}{i,s1}
     \fmf{phantom}{s1,s2}
     \fmf{phantom}{s2,b1}
     \fmf{phantom,tension=0.3}{i,v1}
     \fmf{phantom}{v1,v2}
     \fmf{phantom}{v2,t1}
     \fmf{phantom}{t1,v3}
     \fmf{phantom}{v3,v4}
     \fmf{phantom,tension=0.2}{v4,t2}
\fmffreeze
   \fmf{photon,right}{s1,s2}
   \fmf{photon,left=0.8}{v1,v3}
   \fmf{photon,left=0.8,rubout}{v2,v4}
\end{fmfgraph}}
  \put(7,3){\rule{0.2pt}{28mm}} 
  \put(7,3){\rule{28mm}{0.2pt}} 
  \put(35,3){\rule{0.2pt}{28mm}} 
  \put(7,31){\rule{28mm}{0.2pt}} 
    \put(33,4){\footnotesize 4}
\put(5,1){\rule{0.2pt}{32mm}} 
\put(5,1){\rule{48mm}{0.2pt}} 
\put(53,1){\rule{0.2pt}{32mm}} 
\put(5,33){\rule{48mm}{0.2pt}} 
    \put(51,2){\footnotesize 5}
  \put(16,19){\rule{0.2pt}{10mm}} 
  \put(16,19){\rule{16mm}{0.2pt}} 
  \put(32,19){\rule{0.2pt}{10mm}} 
  \put(16,29){\rule{16mm}{0.2pt}} 
    \put(30,20){\footnotesize 2}
\put(37,11){\rule{0.2pt}{12mm}} 
\put(37,11){\rule{14mm}{0.2pt}} 
\put(51,11){\rule{0.2pt}{12mm}} 
\put(37,23){\rule{14mm}{0.2pt}} 
    \put(49,12){\footnotesize 3}
  \put(13,8){\rule{0.2pt}{9mm}} 
  \put(13,8){\rule{9mm}{0.2pt}} 
  \put(22,8){\rule{0.2pt}{9mm}} 
  \put(13,17){\rule{9mm}{0.2pt}} 
    \put(20,9){\footnotesize 1}
\end{picture}}
\quad = (((s_1)(v_2)v_4)(p_3)v_5)
\end{equation}
(As usual, straight lines stand for fermions and wavy lines for
bosons.) A superficially divergent sector \cite{iz} is necessarily a
region of $\Gamma$ which contains loops. The boxes must be drawn in
such a way that no vertex of $\Gamma$ is situated on the border of the
box and no line of $\Gamma$ is tangential to the border.  Boxes can be
deformed. During the deformation, no vertex is allowed to pass the
border and at no time a line may be tangent to the border of the
box. We consider boxes which differ by a deformation as identical.

We shall work in four dimensional spacetime, but generalization is
obvious. A criterion for superficial divergence of a region confined
in a box is power counting. The box under consideration will contain
$n_B$ bosonic and $n_F$ fermionic external legs. Ghosts are regarded
as bosons here. In a renormalizable theory there can only be a
superficial (ultraviolet) divergence in the box if it contains at
least one loop and if the power counting degree of divergence $d_{pc}$
satisfies
\begin{equation}
d_{pc}:=4-n_B - \tfrac{3}{2} n_F \geq 0~. \label{pc}
\end{equation}
Owing to symmetries the actual degree of divergence $d$ of one graph
or a sum of graphs can be lower than $d_{pc}$ calculated from
\eqref{pc}, see ref.\ \cite{iz}. Examples are graphs in QED with
$n_B=3,\,n_F=0$ (which can be omitted due to Furry's theorem) and with
$n_B=4,\,n_F=0$ (which are superficially convergent due to gauge
symmetry). Always if $d<0$ the box must be erased. This does not mean
that there cannot be divergences in the box to erase. But these
non-superficial divergences must be contained in other boxes which
cannot be deformed into the box we erased.

Our boxes represent the forest structure of $\Gamma$. A forest is a
set of 1PI (one-particle-irreducible, i.e.\ the graph remains
connected after cutting an arbitrary line) divergent subgraphs $\gamma
\subset \Gamma$ which do not overlap.  Instead, any two elements (=
boxes) of a forest are either disjoint or nested. The forest structure
is the collection of the maximal forests of $\Gamma$, i.e.\ the
forests which are not contained in another forest. There are several
maximal forests in general to a Feynman graph.

Kreimer defines \cite{k} a recursive procedure to assign
parenthesized words (PW) to the boxes of a maximal forest. The total
graph $\Gamma$ stands for a certain integrand $I_\Gamma$ depending on
external and internal momenta. A box is represented by a pair of
opening-closing parentheses. Two nested boxes are represented by
$((~)~)$ and two disjoint boxes by $(~)(~)$. In an irreducible PW
(iPW), the leftmost opening parenthesis matches its rightmost closing
parenthesis. A primitive box contains no nested boxes and represents a
graph $\gamma$ without subdivergences. Examples of primitive boxes
$(~)$ are:
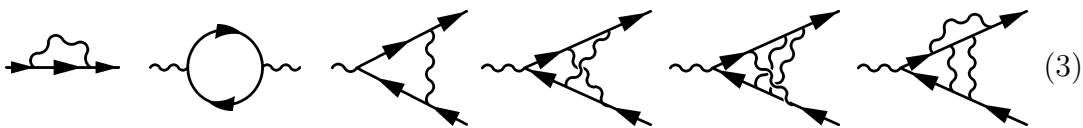
\begin{equation}
\label{prim}
\hskip-3mm
\parbox{15mm}{\begin{fmfgraph}(15,15)
\fmfleft{l}
\fmfright{r}
  \fmf{fermion,tension=2}{l,i1}
  \fmf{fermion}{i1,i2}
  \fmf{fermion,tension=2}{i2,r}
\fmffreeze
  \fmf{photon,left}{i1,i2}
\end{fmfgraph}}
\quad 
\parbox{20mm}{\begin{fmfgraph}(20,15)
\fmfleft{l}
\fmfright{r}
  \fmf{photon,tension=4}{l,i1}
  \fmf{fermion,left}{i1,i2}
  \fmf{fermion,left}{i2,i1}
  \fmf{photon,tension=4}{i2,r}
\end{fmfgraph}}
\quad
\parbox{20mm}{\begin{fmfgraph}(20,15)
\fmfleft{l}
\fmfright{r1,r2}
  \fmf{photon,tension=6}{l,i}
  \fmf{fermion}{i,t1}
  \fmf{fermion,tension=2}{t1,r2}
  \fmf{fermion}{b1,i}
  \fmf{fermion,tension=2}{r1,b1}
\fmffreeze
  \fmf{photon}{t1,b1}
\end{fmfgraph}}
\parbox{25mm}{\begin{fmfgraph}(25,15)
\fmfleft{l}
\fmfright{r1,r2}
  \fmf{photon,tension=2}{l,i}
  \fmf{fermion}{i,t1}
    \fmf{plain}{t1,t2}
  \fmf{fermion}{t2,r2}
  \fmf{fermion}{b1,i}
    \fmf{plain}{b2,b1}
  \fmf{fermion}{r1,b2}
\fmffreeze
  \fmf{photon}{t1,b2}
  \fmf{photon,rubout}{b1,t2}
\end{fmfgraph}}
\parbox{25mm}{\begin{fmfgraph}(25,15)
\fmfleft{l}
\fmfright{r1,r2}
  \fmf{photon,tension=6}{l,i}
  \fmf{phantom}{r1,i}
  \fmf{phantom}{i,r2}
\fmffreeze
  \fmf{fermion}{i,t1}
    \fmf{plain}{t1,t2}
    \fmf{plain,tension=2}{t2,t3}
  \fmf{fermion}{t3,r2}
  \fmf{fermion}{b1,i}
    \fmf{plain,tension=3}{b2,b1}
    \fmf{plain,tension=2}{b3,b2}
  \fmf{fermion}{r1,b3}
\fmffreeze
  \fmf{photon}{b1,t2}
  \fmf{photon}{b2,t3}
  \fmf{photon,rubout}{t1,b3}
\end{fmfgraph}}
\parbox{25mm}{\begin{fmfgraph}(25,15)
\fmfleft{l}
\fmfright{r1,r2}
  \fmf{photon,tension=6}{l,i}
  \fmf{phantom}{r1,i}
  \fmf{phantom}{i,r2}
\fmffreeze
  \fmf{fermion}{i,t1}
    \fmf{plain,tension=2}{t1,t2}
    \fmf{plain,tension=1.5}{t2,t3}
    \fmf{plain,tension=1.5}{t3,t4}
  \fmf{fermion}{t4,r2}
  \fmf{fermion}{b1,i}
    \fmf{plain,tension=3}{b2,b1}
  \fmf{fermion}{r1,b2}
\fmffreeze
  \fmf{photon,left=0.6}{t1,t4}
  \fmf{photon}{b1,t2}
  \fmf{photon}{b2,t3}
\end{fmfgraph}} \hskip-5mm
\end{equation}
(The reader is encouraged to verify using \eqref{pc} that the
last three examples contain no divergent subgraphs.) We associate the
integrand $I_\gamma$ defined by the vertices and propagators of
$\gamma$ to such a primitive box and write the PW $(I_\gamma)$. A
non-primitive box contains nested boxes. It describes a graph $\gamma$
with subdivergences $\gamma_i$, which are already characterized by PWs
$X_i$. Examples for graphs with one nested subdivergence $((~)~)$ are: 
\begin{equation}
\label{nest1}
\parbox{25mm}{\begin{fmfgraph}(25,15)
\fmfleft{l}
\fmfright{r}
  \fmf{fermion}{l,i1}
    \fmf{plain}{i1,i2}
  \fmf{fermion,tension=0.5}{i2,i3}
    \fmf{plain}{i3,i4}
  \fmf{fermion}{i4,r}
\fmffreeze
  \fmf{photon,right=0.7}{i2,i3}
  \fmf{photon,left=0.5}{i1,i4}
\end{fmfgraph}}
\quad 
\parbox{25mm}{\begin{fmfgraph}(25,15)
\fmfleft{l}
\fmfright{r1,r2}
  \fmf{photon,tension=2}{l,i}
  \fmf{fermion}{i,t1}
    \fmf{plain,tension=2}{t1,t2}
  \fmf{fermion}{t2,r2}
  \fmf{fermion}{b1,i}
    \fmf{plain,tension=2}{b2,b1}
  \fmf{fermion}{r1,b2}
\fmffreeze
  \fmf{photon}{t1,b1}
  \fmf{photon}{t2,b2}
\end{fmfgraph}}
\parbox{25mm}{\begin{fmfgraph}(25,15)
\fmfleft{l}
\fmfright{r1,r2}
  \fmf{photon,tension=6}{l,i}
  \fmf{phantom}{r1,i}
  \fmf{phantom}{i,r2}
\fmffreeze
  \fmf{fermion}{i,t1}
    \fmf{plain,tension=1.5}{t1,t2}
    \fmf{plain}{t2,t3}
  \fmf{fermion}{t3,r2}
  \fmf{fermion}{b1,i}
    \fmf{plain,tension=2}{b2,b1}
    \fmf{plain}{b3,b2}
  \fmf{fermion}{r1,b3}
\fmffreeze
  \fmf{photon}{t1,b1}
  \fmf{photon}{b2,t3}
  \fmf{photon,rubout}{b3,t2}
\end{fmfgraph}}
\parbox{25mm}{\begin{fmfgraph}(25,15)
\fmfleft{l}
\fmfright{r1,r2}
  \fmf{photon,tension=6}{l,i}
  \fmf{phantom}{r1,i}
  \fmf{phantom}{i,r2}
\fmffreeze
  \fmf{fermion,tension=0.7}{i,t1}
    \fmf{plain}{t1,t2}
    \fmf{plain}{t2,t3}
  \fmf{fermion}{t3,r2}
  \fmf{fermion,tension=0.7}{b1,i}
    \fmf{plain}{b2,b1}
    \fmf{plain}{b3,b2}
  \fmf{fermion}{r1,b3}
\fmffreeze
  \fmf{photon}{t1,b2}
  \fmf{photon,rubout}{b1,t2}
  \fmf{photon}{b3,t3}
\end{fmfgraph}}
\parbox{25mm}{\begin{fmfgraph}(25,15)
\fmfleft{l}
\fmfright{r1,r2}
  \fmf{photon,tension=6}{l,i}
  \fmf{phantom}{r1,i}
  \fmf{phantom}{i,r2}
\fmffreeze
  \fmf{fermion}{i,t1}
    \fmf{plain}{t1,t2}
    \fmf{plain}{t2,t3}
  \fmf{fermion}{t3,r2}
  \fmf{fermion}{b1,i}
  \fmf{fermion}{r1,b1}
\fmffreeze
  \fmf{photon,left=0.6}{t1,t3}
  \fmf{photon}{b1,t2}
\end{fmfgraph}}
\end{equation}
Examples for graphs with two disjoint nested subdivergences:
$((~)(~)~)$ are:
\begin{equation}
\label{nest2}
\parbox{30mm}{\begin{fmfgraph}(30,15)
\fmfleft{l}
\fmfright{r}
  \fmf{fermion}{l,i1}
    \fmf{plain}{i1,i2}
  \fmf{fermion,tension=0.5}{i2,i3}
    \fmf{plain}{i3,i4}
  \fmf{fermion,tension=0.5}{i4,i5}
    \fmf{plain}{i5,i6}
  \fmf{fermion}{i6,r}
\fmffreeze
  \fmf{photon,right}{i2,i3}
  \fmf{photon,right}{i4,i5}
  \fmf{photon,left=0.5}{i1,i6}
\end{fmfgraph}}
\qquad 
\parbox{25mm}{\begin{fmfgraph}(25,15)
\fmfleft{l}
\fmfright{r1,r2}
  \fmf{photon,tension=6}{l,i}
  \fmf{phantom}{r1,i}
  \fmf{phantom}{i,r2}
\fmffreeze
  \fmf{fermion,tension=0.7}{i,t1}
    \fmf{plain,tension=2}{t1,t2}
    \fmf{plain}{t2,t3}
    \fmf{plain,tension=2}{t3,t4}
  \fmf{fermion}{t4,r2}
  \fmf{fermion,tension=0.7}{b1,i}
    \fmf{plain}{b2,b1}
  \fmf{fermion}{r1,b2}
\fmffreeze
  \fmf{photon,left=0.8}{t2,t3}
  \fmf{photon}{b1,t1}
  \fmf{photon}{b2,t4}
\end{fmfgraph}}
\qquad
\parbox{25mm}{\begin{fmfgraph}(25,15)
\fmfleft{l}
\fmfright{r1,r2}
  \fmf{photon,tension=6}{l,i}
  \fmf{phantom}{r1,i}
  \fmf{phantom}{i,r2}
\fmffreeze
  \fmf{fermion,tension=0.7}{i,t1}
    \fmf{plain,tension=2}{t1,t2}
    \fmf{plain,tension=1.5}{t2,t3}
    \fmf{plain,tension=1.5}{t3,t4}
  \fmf{fermion}{t4,r2}
  \fmf{fermion,tension=0.7}{b1,i}
    \fmf{plain}{b2,b1}
  \fmf{fermion}{r1,b2}
\fmffreeze
  \fmf{photon,left=0.6}{t2,t4}
  \fmf{photon}{b1,t1}
  \fmf{photon}{b2,t3}
\end{fmfgraph}}
\end{equation}
\iflowmem\end{fmffile}\begin{fmffile}{fmfhopf2}\fi%
And here are two examples for graphs with a nested subdivergence which
has itself a nested subsubdivergence $(((~)~)~)$:
\begin{equation}
\label{nestsub}
\parbox{35mm}{\begin{fmfgraph}(35,15)
\fmfleft{l}
\fmfright{r}
  \fmf{fermion}{l,i1}
    \fmf{plain,tension=2}{i1,i2}
  \fmf{fermion}{i2,i3}
    \fmf{plain}{i3,i4}
  \fmf{fermion}{i4,i5}
    \fmf{plain,tension=2}{i5,i6}
  \fmf{fermion}{i6,r}
\fmffreeze
  \fmf{photon,left}{i3,i4}
  \fmf{photon,right=0.7}{i2,i5}
  \fmf{photon,left=0.5}{i1,i6}
\end{fmfgraph}}
\qquad 
\qquad 
\parbox{25mm}{\begin{fmfgraph}(25,15)
\fmfleft{l}
\fmfright{r1,r2}
  \fmf{photon,tension=2}{l,i}
  \fmf{fermion,tension=0.5}{i,t1}
    \fmf{plain}{t1,t2}
    \fmf{plain}{t2,t3}
  \fmf{fermion}{t3,r2}
  \fmf{fermion,tension=0.5}{b1,i}
    \fmf{plain}{b2,b1}
    \fmf{plain}{b3,b2}
  \fmf{fermion}{r1,b3}
\fmffreeze
  \fmf{photon}{t1,b1}
  \fmf{photon}{t2,b2}
  \fmf{photon}{t3,b3}
\end{fmfgraph}}
\end{equation}
If we shrink all nested boxes (=divergent subgraphs $\gamma_i$) of
$\gamma$ to points, there remains a fraction $I_{\gamma/\cup\gamma_i}$
of the integrand of $\gamma$ defined by the vertices and propagators
of $\gamma/ \cup \gamma_i$. The latter should be regarded as a Feynman
graph with holes at the places where the subgraphs $\gamma_i$ had been
before. We agree that for self-energy insertions $\gamma_i$ splitting
propagators into two, one of the new propagators belongs to the
subgraph $\gamma_i$. In this way we keep the number of possible holes
in a Feynman graph finite. We write the fraction
$I_{\gamma/\cup\gamma_i}$ next to the right closing parenthesis and
everything we have shrunk to a point (the $X_i$) between that fraction
and the left opening parenthesis. The resulting PW looks like this:
$\big(X_1\dots X_n\; I_{\gamma/\{\gamma_1\cup \dots \cup \gamma_n\}}
\big)$. Note that the order of disjoint boxes is irrelevant. For
instance, the PW of the example \eqref{bsp} (considered as 1PI) looks
as follows:
\begin{align}
&(((s_1)(v_2)v_4)(p_3)v_5) \\
&= \left(\left(\left(
\parbox{17mm}{\begin{fmfgraph*}(17,15)
\fmfleft{l}
\fmfright{r}
  \fmf{plain}{l,i1}
  \fmf{fermion,tension=0.5}{i1,i2}
  \fmf{plain}{i2,r}
\fmffreeze
  \fmf{photon,left}{i1,i2}
\fmfv{label=/,label.angle=180,label.dist=20}{i1}
\end{fmfgraph*}}
\right)
\left(
\parbox{20mm}{\begin{fmfgraph*}(20,15)
\fmfleft{l}
\fmfright{r1,r2}
  \fmf{photon,tension=2}{l,i}
  \fmf{fermion}{i,t1}
    \fmf{plain}{t1,t2}
  \fmf{plain}{t2,r2}
  \fmf{plain}{b1,i}
    \fmf{plain}{b2,b1}
  \fmf{plain}{r1,b2}
\fmffreeze
  \fmf{photon}{t1,b2}
  \fmf{photon,rubout}{b1,t2}
\fmfv{label=$\backslash$,label.angle=19,label.dist=20}{t2}
\fmfv{label=/,label.angle=-20,label.dist=20}{b2}
\fmfv{label=/,label.angle=180,label.dist=20}{i}
\end{fmfgraph*}}
\right)
\parbox{20mm}{\begin{fmfgraph*}(20,15)
\fmfleft{l}
\fmfright{r1,r2}
  \fmf{photon,tension=4}{l,i}
  \fmf{fermion}{i,t1}
  \fmf{plain,tension=2}{t1,r2}
  \fmf{phantom}{b1,i}
  \fmf{plain,tension=2}{r1,b1}
\fmffreeze
  \fmf{photon}{t1,b1}
  \fmf{plain}{b1,b0}
  \fmf{plain}{b0,i}
\fmfv{decor.shape=circle,decor.filled=empty,decor.size=2mm,%
      label=$\backslash$,label.angle=19,label.dist=20}{t1}
\fmfv{decor.shape=circle,decor.filled=empty,decor.size=2mm,%
      label=/,label.angle=-20,label.dist=19}{b0}
\fmfv{label=/,label.angle=-180,label.dist=20}{i}
\fmfv{label=/,label.angle=-20,label.dist=20}{b1}
\end{fmfgraph*}}
\right)\left(
\parbox{17mm}{\begin{fmfgraph*}(17,15)
\fmfleft{l}
\fmfright{r}
  \fmf{photon,tension=4}{l,i1}
  \fmf{fermion,left}{i1,i2}
  \fmf{plain,left}{i2,i1}
  \fmf{photon,tension=4}{i2,r}
\fmfv{label=$/$,label.angle=-180,label.dist=20}{i1}
\end{fmfgraph*}}
\right)
\parbox{20mm}{\begin{fmfgraph*}(20,15)
\fmfleft{l}
\fmfright{r1,r2}
  \fmf{photon,tension=4}{l,i}
  \fmf{fermion}{i,t1}
  \fmf{plain,tension=2}{t1,r2}
  \fmf{plain}{b1,i}
  \fmf{plain,tension=2}{r1,b1}
\fmffreeze
  \fmf{photon}{t1,i1}
  \fmf{photon}{i1,b1}
\fmfv{decor.shape=circle,decor.filled=empty,decor.size=2mm,%
      label=/,label.angle=180,label.dist=20}{i}
\fmfv{decor.shape=circle,decor.filled=empty,decor.size=2mm,%
      label=\mbox{---},label.angle=-90,label.dist=18}{i1}
\fmfv{label=$\backslash$,label.angle=19,label.dist=20}{t1}
\fmfv{label=/,label.angle=-20,label.dist=20}{b1}
\end{fmfgraph*}}
\right) \notag
\end{align}
A slash through a propagator means amputation and a small circle
symbolizes a hole. We see that our building blocks are the Feynman
graphs with possible holes at any vertex and in any propagator.

By this procedure we associate a PW to each maximal forest. As
discovered by Kreimer \cite{k}, the PWs form a Hopf algebra whose
antipode axiom reproduces the forest formula \cite{z}. This assumes
that overlapping divergences such as 
\begin{equation}
\parbox{24mm}{\begin{fmfgraph}(24,12)
\fmfleft{l} 
\fmfright{r} 
\fmfbottom{b} 
\fmftop{t} 
  \fmf{photon,tension=2}{l,l1}
  \fmf{photon,tension=2}{r,r1}
    \fmf{fermion,left=.4}{l1,t}
    \fmf{fermion,left=.4}{b,l1}
    \fmf{fermion,left=.4}{t,r1}
    \fmf{fermion,left=.4}{r1,b}
\fmffreeze
  \fmf{photon}{t,b}
\end{fmfgraph}}
\qquad\qquad
\parbox{30mm}{\begin{fmfgraph}(30,15)
\fmfleft{l}
\fmfright{r}
  \fmf{fermion}{l,i1}
    \fmf{plain}{i1,i2}
  \fmf{fermion,tension=0.5}{i2,i3}
    \fmf{plain}{i3,i4}
  \fmf{fermion}{i4,r}
\fmffreeze
  \fmf{photon,left=0.7}{i1,i3}
  \fmf{photon,right=0.7}{i2,i4}
\end{fmfgraph}}
\end{equation}
\iflowmem\end{fmffile}\begin{fmffile}{fmfhopf3}\fi%
have been disentangled into a linear combination of PWs containing
disjoint and nested divergences exclusively, for instance via the
Schwinger-Dyson equation, see \cite{k,kh}. The outcome is thus a
linear combination of PWs each of them describing a maximal forest,
and the forest formula is reduced to a rather trivial prescription.

The goal of this paper is to modify the PWs and the Hopf algebra
operations in such a way that any 1PI-Feynman graph is described by
\emph{a single} PW and that all Hopf algebra operations are defined on
such a PW. Our starting point is the observation that in the case of
overlapping divergences there exist several maximal forests to a
Feynman graph. It is clear that democracy requires to comprise all PWs
associated to these maximal forests to one bigger object. We propose
to build a column vector whose components are the PWs of maximal
forests. The order of the components (or rows as they are long
objects) of this vector is not relevant, of course.  As the integrands
associated to the PWs of each row are equal, we associate this
universal integrand to our column vector.

There is one further modification necessary. Later on we are going to
identify the subwords of such a vector and define the removal of
subwords. Subwords represent subgraphs and the removal means replacing
the subgraph by a hole. But subgraphs or subwords can occur
identically in various maximal forests. If we now compare the maximal
forests of a graph with removed subgraph and the maximal forests of
the original graph, it is easy to see that the subgraph is removed in
all maximal forest it had occurred. (An example is the step from
\eqref{PVV} to \eqref{PV} in the next section by cutting out loop
$3$.)  We must implement this feature in our vectors. We propose to
connect by a tree of lines the closing parentheses of identical and
simultaneously shrinkable boxes. If we pull out a subword of such a
vector and if the subword is connected over various rows, we simply
have to remove all of them.

Thus, our PWs are vectors of one-line-PWs representing the maximal
forests of a Feynman graph, where the closing parentheses of
simultaneously shrinkable boxes are connected. We define now the
notion of a parenthesized subword (PSW) of a PW.  A PSW $Y$ of 
$X$ is everything between a set of connected closing parentheses and
its matching opening parentheses. Disconnected rows of $X$ which are
accidentally between connected rows are not part of the PSW $Y$ under
consideration.

There is an algorithm which yields all PSW of a PW. Starting with the
first row we run from the left through the PW until we meet a closing
parenthesis. In general, it will be connected with other closing
parentheses in different rows. These connected closing parentheses and
their matching opening parentheses define our first PSW.  We mark all
these connected closing parentheses. We then go ahead and move through
the first row until we arrive at the next closing parenthesis. This
gives the next PSW and marks the next set of parentheses. We repeat
this procedure until the rightmost closing parenthesis is
reached. Then we pass to the second row and continue to search for 
new closing parentheses and related PSW, i.e.\ we ignore all
parentheses marked in the previous steps. This search continues
through all rows and stops at the lower right corner of our PW.

In what follows we will freely use the notions parenthesized word
(PW), irreducible PW (iPW, the leftmost and rightmost parentheses
match), primitive PW (no nested divergences, a special iPW) and
parenthesized subword (PSW, a special iPW). We remark that a possible
extension could be the inclusion of superficially convergent
1PI-graphs ($d < 0$) with subdivergences. All finite integrands fuse
and stand immediately before the rightmost closing parentheses.

We will give now some examples for Feynman graphs with overlapping
divergences which are represented by parenthesized words of several
maximal forests. The PSW of some of these examples are discussed and
further evaluated in section~\ref{ex2}.

\section{Examples for Feynman graphs with several maximal forests}
\label{ex1}

In QED there is the following contribution to the photon propagator:
\begin{equation}
\parbox{36mm}{\begin{fmfgraph*}(36,18)
\fmfleft{l} 
\fmfright{r} 
\fmfbottom{b} 
\fmftop{t} 
  \fmf{photon,tension=2}{l,l1}
  \fmf{photon,tension=2}{r,r1}
    \fmf{fermion,left=.4,label=\mbox{\small$k_1{+}p$},%
                 label.dist=10}{l1,t}
    \fmf{fermion,left=.4,label=\mbox{\small$k_1$},%
                 label.dist=6}{b,l1}
    \fmf{fermion,left=.4,label=\mbox{\small$k_2{+}p$},%
                 label.dist=10}{t,r1}
    \fmf{fermion,left=.4,label=\mbox{\small$k_2$},%
                 label.dist=10}{r1,b}
\fmffreeze
  \fmf{photon}{t,b}
  \fmf{phantom_arrow}{t,b}
  \fmf{phantom_arrow,label=\mbox{\small$p$}}{l,l1}
  \fmf{phantom_arrow,label=\mbox{\small$p$}}{r1,r}
  \fmfv{label=\mbox{\small$\kappa$},label.dist=7,label.angle=90}{t}
  \fmfv{label=\mbox{\small$\kappa$},label.dist=9,label.angle=-90}{b}
  \fmfv{label=\mbox{\small$\mu$},label.dist=13,label.angle=-25}{l1}
  \fmfv{label=\mbox{\small$\nu$},label.dist=10,label.angle=-150}{r1}
\fmffreeze
  \fmf{phantom}{t,c}
  \fmf{phantom}{c,b}
\fmffreeze
  \fmfv{label=\mbox{\small$\hskip 2pt k_1-\!k_2 \hskip -2pt$},%
        label.dist=18,label.angle=68}{c}
\end{fmfgraph*}}
\rule[-10.5mm]{0pt}{23mm}
\qquad \qquad
\begin{array}{rl} 
((v_1) p_2) &\VL \\ 
((v_2) p_1) &\AL
\end{array} 
\label{PV}
\end{equation}
We can draw two maximal forests of boxes. We can first draw a box
around the left loop which contains the vertex correction with interior
momentum $k_1$. Then we put this box into the large box which
encircles both loops. Or we can first enclose the right loop by a
vertex box and then put everything into \emph{the same} large
box. Graphically, the two possibilities look like this:
\begin{equation}
\parbox{31mm}{
\begin{picture}(31,24)
\put(0,11.5){\parbox{30mm}{\begin{fmfgraph*}(30,15)
  \fmfkeep{polar}
\fmfleft{l} 
\fmfright{r} 
\fmfbottom{b} 
\fmftop{t} 
  \fmf{photon,tension=2}{l,l1}
  \fmf{photon,tension=2}{r,r1}
    \fmf{fermion,left=.4}{l1,t}
    \fmf{fermion,left=.4}{b,l1}
    \fmf{fermion,left=.4}{t,r1}
    \fmf{fermion,left=.4}{r1,b}
       \fmfv{label=$1$,label.dist=25,label.angle=0}{l1}
       \fmfv{label=$2$,label.dist=25,label.angle=180}{r1}
\fmffreeze
  \fmf{photon}{t,b}
\end{fmfgraph*}}}
\put(6,3){\rule{0.2pt}{19mm}} 
\put(6,3){\rule{11mm}{0.2pt}} 
\put(17,3){\rule{0.2pt}{19mm}} 
\put(6,22){\rule{11mm}{0.2pt}} 
  \put(4,1){\rule{0.2pt}{23mm}} 
  \put(4,1){\rule{21mm}{0.2pt}} 
  \put(25,1){\rule{0.2pt}{23mm}} 
  \put(4,24){\rule{21mm}{0.2pt}} 
\end{picture}} = ((v_1)p_2)
\qquad 
\mbox{or}
\qquad
\parbox{31mm}{
\begin{picture}(31,24)
\put(0,11.5){\parbox{30mm}{\fmfreuse{polar}}}
\put(13,3){\rule{0.2pt}{19mm}} 
\put(13,3){\rule{11mm}{0.2pt}} 
\put(24,3){\rule{0.2pt}{19mm}} 
\put(13,22){\rule{11mm}{0.2pt}} 
  \put(5,1){\rule{0.2pt}{23mm}} 
  \put(5,1){\rule{21mm}{0.2pt}} 
  \put(26,1){\rule{0.2pt}{23mm}} 
  \put(5,24){\rule{21mm}{0.2pt}} 
\end{picture}} 
= ((v_2)p_1)~.
\label{PV0}
\end{equation}
In the first case, the innermost box is the primitive box $(v_1)$
the integrand of which is -- in the Feynman gauge -- given by
\[
v^{\mu B}_{1A} 
= \left[ e  \gamma^\kappa \, \tfrac{\kslash_1 
+\mu}{k_1^2-\mu^2} \, e \gamma^\mu \, 
\tfrac{\kslash_1 + \pslash +\mu}{(k_1+p)^2-\mu^2} \,
e \gamma_\kappa \, \tfrac{1}{(k_1-k_2)^2-M^2} \right]_A^B ~= 
\quad
\parbox{20mm}{\begin{fmfgraph*}(20,15)
\fmfleft{l}
\fmfright{r1,r2}
  \fmf{photon,tension=3,label=\mbox{\small$\mu~$},label.dist=14}{l,i}
  \fmf{plain,tension=2,label=\mbox{\small$~A$},label.dist=7,%
       label.side=right}{r1,b}
  \fmf{plain}{b,i}
  \fmf{fermion}{i,t}
  \fmf{plain,tension=2,label=\mbox{\small$~B$},label.dist=7,%
       label.side=right}{t,r2}
\fmffreeze
  \fmf{photon,label=\mbox{\small$1$},label.side=right}{t,b}
\fmfv{label=/,label.angle=180,label.dist=20}{i}
\fmfv{label=$\backslash$,label.angle=19,label.dist=20}{t}
\fmfv{label=/,label.angle=-20,label.dist=20}{b}
\end{fmfgraph*}} 
\]
Here, $e$ is the electron charge, $\mu$ is the electron mass and $M$
an auxiliary photon mass to avoid IR-divergences. Capital roman
letters label Clifford indices and greek letters Lorentz indices. This
vertex box is nested in the large box, so we must write $((v_1)p_2)$
as the maximal forest. The integrand $p_2$ is the interior of the
large box after shrinking the small box $(v_1)$ to a hole. What
remains is loop $2$ and the integrand is found to be
\[
p^{\nu A}_{2B} 
= \left[ \tfrac{\kslash_2+\pslash +\mu}{(k_2+p)^2-\mu^2} 
\, e \gamma^\nu \, 
\tfrac{\kslash_2 + \mu}{k_2^2-\mu^2} \right]_B^A ~=
\quad
\parbox{25mm}{\begin{fmfgraph*}(25,20)
\fmfleft{l}
\fmfright{r}
  \fmf{photon,tension=3}{l,i1}
  \fmf{photon,tension=3,label=\mbox{\small$~\nu$},label.dist=14,%
       label.side=right}{i2,r}
  \fmf{plain,left}{i2,i1}
  \fmf{fermion,left}{i1,i2}
\fmfv{label=\mbox{{\small$2$}\hskip 7mm /\hskip-4mm},%
      label.angle=180,label.dist=0.1pt}{i2}
\fmfv{decor.shape=circle,decor.filled=empty,decor.size=2mm,%
      label=\mbox{/\hskip 1mm \small%
      $\begin{array}{c}B \\[3mm] A\end{array}$},%
      label.angle=180,label.dist=0.1pt}{i1}
\end{fmfgraph*}} 
\]
In the second case the loops $1$ and $2$ change their role and we
obtain the maximal forest $((v_2)p_1)$ with
\begin{align*}
v^{\nu A}_{2B} &= \left[ e\gamma_\kappa  \,
\tfrac{\kslash_2 +\pslash+\mu}{(k_2+p)^2-\mu^2} 
\, e \gamma^\nu \, \tfrac{\kslash_2 +\mu}{k_2^2-\mu^2}
\, e\gamma^\kappa \, \tfrac{1}{(k_2-k_1)^2-M^2} \right]^A_B~,
\\
p^{\mu B}_{1A} &= \left[ \tfrac{\kslash_1 +\mu}{k_1^2-\mu^2} 
\, e \gamma^\mu \, \tfrac{\kslash_1 + \pslash +\mu}{
(k_1+p)^2-\mu^2} \right]^B_A ~.
\end{align*}
We have found two maximal forests $((v_1)p_2)$ and $((v_2)p_1)$ in
this example. These two forests form the 2-line vector 
$\begin{array}{l} ((v_1)p_2) \rule{0pt}{1.8ex} \\[-0.5ex] ((v_2)p_1) 
\rule[-0.5ex]{0pt}{1.5ex} \end{array}$. However, the large box
occurs identically in both maximal forests. We cannot shrink it in one
of them and keep it in the other. Therefore, the closing parentheses
representing the large box in both rows of the vector must be connected,
as we have already indicated in \eqref{PV}.

Here is a graph with two maximal forests containing a nested
divergence:
\begin{equation}
\parbox{40mm}{\rule{0pt}{20mm}\begin{fmfgraph*}(30,15)
\fmfkeep{fmfPVV}
\fmfleft{l} 
\fmfright{r} 
\fmfbottom{b} 
\fmftop{t1,t2,t,t3,t4} 
  \fmf{photon,tension=2}{l,l1}
  \fmf{photon,tension=2}{r,r1}
    \fmf{phantom,left=.4}{l1,t}
    \fmf{fermion,left=.4}{b,l1}
    \fmf{phantom,left=.4}{t,r1}
    \fmf{fermion,left=.4}{r1,b}
       \fmfv{label=$1$,label.dist=35,label.angle=0}{l1}
       \fmfv{label=$2$,label.dist=35,label.angle=180}{r1}
\fmffreeze
  \fmf{photon}{t,b}
\fmffreeze
     \fmf{plain,left=.18}{t,r2}
     \fmf{fermion,left=.2,tension=0.5}{r2,r1}
  \fmf{phantom,tension=1.2}{t3,r2}
\fmffreeze
     \fmf{fermion,left=.2,tension=0.5}{l1,l2}
     \fmf{plain,left=.18}{l2,t}
  \fmf{phantom,tension=1.2}{t2,l2}
\fmffreeze
  \fmf{photon,left,tension=0.5}{l2,r2}
       \fmfv{label=\mbox{\small$3$},label.dist=7,label.angle=64}{t}
\end{fmfgraph*}}
\qquad
\begin{array}{rrrlrrl} 
(((v_3) &\VL& v_{13}) p_2) &\VL \\ 
(((v_3) &\AL& v_{23}) p_1) &\AL
\end{array} 
\label{PVV}
\end{equation}
\iflowmem\end{fmffile}\begin{fmffile}{fmfhopf4}\fi%
The vertex correction $v_3$ is nested in both vertex corrections
$v_{i3}$ comprising the common loop $3$ and loop $i$. The subword
$(v_3)$ is identical in both maximal forests $(((v_3)v_{13})p_2)$ and
$(((v_3)(v_{23})p_1)$. If we shrink it in one of them it is
automatically removed in the other one. For the same
reasons both maximal forests are connected at the outermost box.

Here is now a more complicated forest structure:
\begin{equation}
\parbox{40mm}{\begin{fmfgraph*}(30,15)
\fmfleft{l} 
\fmfright{r} 
\fmfbottom{b1,b2,b3,b4,b5,b6} 
\fmftop{t1,t2,t3,t4,t5,t6} 
  \fmf{photon,tension=2}{l,l1}
  \fmf{photon,tension=2}{r,r1}
    \fmf{fermion,left=.35}{l1,t3}
    \fmf{fermion,left=.35}{b3,l1}
      \fmf{plain,left=.1}{t3,t4}
      \fmf{plain,right=.1}{b3,b4}
    \fmf{fermion,left=.35}{t4,r1}
    \fmf{fermion,left=.35}{r1,b4}
       \fmfv{label=$1$,label.dist=20,label.angle=0}{l1}
       \fmfv{label=$2$,label.dist=20,label.angle=180}{r1}
\fmffreeze
  \fmf{photon}{t3,b3}
  \fmf{photon,label=$3$,label.side=right}{t4,b4}
\end{fmfgraph*}}
\qquad
\begin{array}{lrrrr} 
( &(v_1) \VL&(v_2) \VL&         & p_3) \VL \\ 
((&(v_1) \AL&      \IL& v_{13}) & p_2) \TL \\
((&         &(v_2) \AL& v_{23}) & p_1) \AL
\end{array} 
\label{VVP}
\end{equation}
We have three possibilities for drawing disjoint boxes: We can take
loops $1$ and $2$ and put them into the large box, or we can put loop
$1$ into the vertex box which covers loops $1$ and $3$ and then
everything into the large box, or we can exchange the role of loops
$1$ and $2$. 

Let us also give an example from $\phi^4$-theory. There is the
following second-order correction to the propagator:
\begin{equation}
\parbox{40mm}{\begin{fmfgraph*}(30,15)
  \fmfleft{i} 
  \fmf{photon,tension=5}{i,j1}
  \fmf{photon,left,label=$1~~{}$,label.dist=5}{j1,j2}
  \fmf{photon,right,label=${}~~3$,label.dist=5}{j1,j2}
  \fmf{photon,label=${}~2$,label.dist=5}{j1,j2}
  \fmf{photon,tension=5}{j2,f}
  \fmfright{f} 
\end{fmfgraph*}}
\qquad
\begin{array}{rrr} ((x_{23}) & y_1) & \VL \\ 
((x_{31}) & y_2) & \TL \\ ((x_{12}) & y_3) & \AL 
\end{array}
\label{phi4}
\end{equation}
Here, $x_{ij}$ is the vertex correction 
\begin{fmfgraph}(10,3) 
\fmfleft{l1,l2}
\fmfright{r1,r2}
\fmf{plain,right=0.1,tension=3}{l2,i1}
\fmf{plain,right=0.4}{i1,i2}
\fmf{plain,right=0.1,tension=3}{i2,r2}
\fmf{plain,left=0.1,tension=3}{l1,i1}
\fmf{plain,left=0.4}{i1,i2}
\fmf{plain,left=0.1,tension=3}{i2,r1}
\end{fmfgraph}
involving the lines $i,j$ and $y_k$ the tadpole graph 
\begin{fmfgraph}(10,3)
\fmfleft{l,l1}
\fmfright{r,r1}
\fmftop{t}
\fmf{plain}{l,i}
\fmf{plain}{i,r}
\fmffreeze
\fmf{plain,left}{i,t}
\fmf{plain,left}{t,i}
\end{fmfgraph}
\iflowmem\end{fmffile}\begin{fmffile}{fmfhopf5}\fi%
involving the line $k$. The three maximal forests are connected
because shrinking one of them to a hole forces the reduction of the
other two.

\section{Kreimer's $R$-operation \normalfont{\protect\cite{k}}}
\label{ro}

To any PW $X$, Kreimer associates a second, in a certain sense
equivalent copy $R[X]$. The philosophy is that $R[X]$ is a local
counterterm, a point-like interaction. It is so to say a new vertex,
mass or kinetic term in the Lagrangian, which itself is infinite but
such that a certain combination of counterterms and divergent 1PI
graphs is finite. The finite linear combination in question is given
by the forest formula or~-- as discovered by Kreimer~-- by the
antipode axiom of a (quasi-) Hopf algebra to construct. For
renormalizability it is essential that all counterterms can be
absorbed by a redefinition of physical parameters of the theory.  In
particular in gauge theories there are potentially more types of
counterterms than physical parameters \cite{iz}. It is important then
that counterterms and divergences of the sum of all graphs
contributing to a certain amplitude cancel. We avoid a discussion of
these subtleties by considering scalar theories or~-- with some
care~-- QED.

The $R$-operation depends on the renormalization scheme, which in
principle is arbitrary but fixed throughout the investigation. We
shall work in the BPHZ scheme \cite{bp,h,z} which is the standard one
in connection with the forest formula. A iPW $X$ represents one box
containing a divergent Feynman graph with in general several forests
of subdivergences. The box has $n_B$ bosonic and $n_F$ fermionic
external legs. The superficial degree of divergence $d[X]$ of the iPW
$X$ is bounded by the power counting theorem \eqref{pc}, $d[X] \leq 4
{-} n_B {-} \tfrac{3}{2} n_F$. In the BPHZ scheme the integrand $R[X]$
is the Taylor expansion until order $d[X]$ with respect to the
external momenta of $X$. We call $X=\prod_i X_i$ a tree if each $X_j
\subset X$ has a common momentum variable with at least one $X_i
\subset X$, $i\neq j$. In this case we define $R[X]$ to be the Taylor
expansion with respect to the external momenta of the smallest
possible iPW $\tilde{X}$ containing all $X_i$ as subwords. Finally,
for $X$ being a product of disjoint trees $X_t$, we define $R[\prod
X_t]=\prod R[X_t]$. Note that in general $X-R[X]$ is an integrand
yielding a finite integral only if $X$ is a primitive PW without
subdivergences.

To give an example, consider the divergent Feynman graph with
subdivergence 
\begin{align}
&\parbox{57mm}{\begin{fmfgraph*}(57,30)
\fmfleft{l}
\fmfright{r1,r2}
\fmf{photon,tension=3,label=\mbox{\small$p_1{-}p_2$},%
            label.side=right,label.dist=10}{l,i}
\fmf{phantom_arrow,tension=0}{l,i}
  \fmf{fermion,tension=1.5,label=\parbox{30mm}{%
       \small$p_1{-}p_2{+}k_1$},label.side=left,label.dist=15}{i,t1} 
  \fmf{fermion,tension=1.5,label=\mbox{\small$k_1$},%
               label.side=left,label.dist=8}{b1,i}
  \fmf{fermion,tension=1.3,label=\parbox{30mm}{%
       \small$p_1{-}p_2{+}k_2$},label.side=left,label.dist=15}{t1,t2}
  \fmf{fermion,tension=1.3,label=\mbox{\small$k_2$},%
               label.side=left,label.dist=10}{b2,b1}
  \fmf{fermion,tension=2,label=\mbox{\small$p_1$},%
               label.side=left,label.dist=15}{t2,r2}
  \fmf{fermion,tension=2,label=\mbox{\small$p_2$},%
               label.side=left,label.dist=15}{r1,b2}
\fmffreeze
\fmf{photon,label=\mbox{\small$k_1{-}k_2$},%
            label.side=left,label.dist=13}{t1,b1}
\fmf{phantom_arrow,tension=0}{t1,b1}
\fmf{photon,label=\mbox{\small$k_2{-}p_2$},%
            label.side=left,label.dist=13}{t2,b2}
\fmf{phantom_arrow,tension=0}{t2,b2}
\fmfv{label=\mbox{\small$\mu$},label.dist=3,label.angle=105}{i}
\fmfv{label=\mbox{\small$\nu$},label.dist=10,label.angle=-30}{t1}
\fmfv{label=\mbox{\small$\nu$},label.dist=7,label.angle=45}{b1}
\fmfv{label=\mbox{\small$\kappa$},label.dist=10,label.angle=-30}{t2}
\fmfv{label=\mbox{\small$\kappa$},label.dist=7,label.angle=45}{b2}
\end{fmfgraph*}}
\!\!\! = 
\Bigg( \! \Bigg( \parbox{20mm}{\begin{fmfgraph*}(20,15)
\fmfleft{l}
\fmfright{r1,r2}
  \fmf{photon,tension=4,label=\mbox{\small$\;\mu$},label.dist=14,%
       label.side=right}{l,i} 
  \fmf{fermion}{i,t1}
  \fmf{plain,tension=2,label=\mbox{\small$~C$},label.dist=7,%
       label.side=right}{t1,r2}
  \fmf{plain}{b1,i}
  \fmf{plain,tension=2,label=\mbox{\small$~B$},label.dist=7,%
       label.side=right}{r1,b1}
\fmffreeze
  \fmf{photon}{t1,b1}
\fmfv{label=$\backslash$,label.angle=19,label.dist=20}{t1}
\fmfv{label=/,label.angle=-180,label.dist=20}{i}
\fmfv{label=/,label.angle=-20,label.dist=20}{b1}
\end{fmfgraph*}}\;
\Bigg) 
\parbox{20mm}{\begin{fmfgraph*}(20,15)
\fmfleft{l}
\fmfright{r1,r2}
  \fmf{photon,tension=4}{l,i}
  \fmf{fermion}{i,t1}
  \fmf{plain,tension=2,label=\mbox{\small$~D$},label.dist=7,%
       label.side=right}{t1,r2}
  \fmf{plain}{b1,i}
  \fmf{plain,tension=2,label=\mbox{\small$~A$},label.dist=7,%
       label.side=right}{r1,b1}
\fmffreeze
  \fmf{photon}{t1,b1}
\fmfv{label=$\backslash$,label.angle=19,label.dist=20}{t1}
\fmfv{decor.shape=circle,decor.filled=empty,decor.size=2mm,%
      label=/,label.angle=-180,label.dist=20}{i}
\fmfv{label=/,label.angle=-20,label.dist=20}{b1}
\fmfv{label=\mbox{\small$\begin{array}{c}C \\[1.5mm] 
      \raisebox{-0.5mm}{B} \end{array}$},
      label.dist=3mm,label.angle=0}{l}
\end{fmfgraph*}}\;
\Bigg)  = ((v_1)v_2)~,
\\*[1ex]
&v_{1B}^{\mu C} = \left[ e\gamma^\nu \,
\tfrac{\kslash_1 + \mu}{k_1^2-\mu^2} \, e\gamma^\mu \,
\tfrac{\kslash_1 +(\pslash_1-\pslash_2 +\kslash_2)
-\kslash_2 +\mu}{(k_1+(p_1-p_2+k_2)-k_2)^2-\mu^2} \,
e \gamma_\nu \,\tfrac{1}{(k_1-k_2)^2-M^2} \right]_B^C \;, \notag 
\\
&v_{2AC}^{BD} = \left[ e \gamma^\kappa \, 
\tfrac{\kslash_2 + \mu}{k_2^2 -\mu^2} \right]_A^B
\left[\tfrac{\pslash_1-\pslash_2+\kslash_2 +\mu}{
(p_1-p_2+k)^2 -\mu^2} \, e \gamma_\kappa \, 
\tfrac{1}{(k_2-p_2)^2-M^2} \right]_C^D~.
\notag
\end{align}
\iflowmem\end{fmffile}\begin{fmffile}{fmfhopf6}\fi%
We have written $v_1$ in a form where its external momenta
$p_1{-}p_2{+}k_2$ and $k_2$ are explicit. The two subwords of
$((v_1)v_2)$ are clearly $(v_1)$ and $((v_1)v_2)$. Let us compute
$R[(v_1)]$. It has $2$ fermionic and $1$ bosonic external legs, hence
$d[(v_1)]\leq 0$, and actually $d[(v_1)] = 0$. In the BPHZ scheme we
take the Taylor expansion of $(v_1)$ in its external momenta
$p_1{-}p_2{+}k_2$ and $k_2$ until order $0$. This gives
\begin{align}
R[(v_1)] &= v_{1B}^{\mu C} \big|_{p_1-p_2+k_2=k_2=0}
=\left[ e\gamma^\nu \, \tfrac{\kslash_1 + \mu}{k_1^2-\mu^2} \,
e\gamma^\mu \, \tfrac{\kslash_1 + \mu}{k_1^2-\mu^2} \,
e \gamma_\nu\,\tfrac{1}{k_1^2-M^2} \right]_B^C \notag 
\\*[1ex]
& = ~~ \parbox{30mm}{\begin{fmfgraph*}(30,15)
\fmfleft{l}
\fmfright{r1,r2}
  \fmf{photon,label=\mbox{\small$p_1{-}p_2;\mu$},%
               label.side=right,label.dist=14}{l,i}
  \fmf{fermion,label=\mbox{\small$p_1{-}p_2{+}k_2;C$},%
               label.side=left,label.dist=10}{i,r2}
  \fmf{fermion,label=\mbox{\small$k_2;B$},%
               label.side=left,label.dist=10}{r1,i}
\fmfdot{i}
\fmfv{label=\mbox{/\hskip 5mm $\begin{array}{c} \backslash \\ /
      \end{array}\hskip -4mm$},label.dist=1,label.angle=180}{i}
\fmffreeze
\fmf{phantom_arrow}{l,i}
\fmfkeep{Rv1}
\end{fmfgraph*}}
\label{VVb}
\end{align}
We see that $R[(v_1)]$ defines a local counterterm, and the integral
$\int d^4 k_1 \;\{ (v_1)-R[(v_1)]\}$ is finite.

We can now apply the $R$-operation to the PWs $((v_1)v_2)$ and
$R[(v_1)] (v_2)$. In both cases this means Taylor expansion with
respect to the external momenta $p_1,p_2$ of $((v_1)v_2)$ until degree
$d[((v_1)v_2)]=0$, because $R[(v_1)]$ and $(v_2)$ have common momenta
$p_1,p_2,k_2$. We obtain
\begin{subequations}
\begin{align}
R[ & R[(v_1)] (v_2)] = ~ \parbox{30mm}{\fmfreuse{Rv1}}  \times~
\parbox{30mm}{\begin{fmfgraph*}(30,15)
\fmfleft{l}
\fmfright{r1,r2}
  \fmf{photon,label=\mbox{\small$p_1{-}p_2~~{}$},%
               label.side=right,label.dist=14}{l,i}
  \fmf{fermion,label=\mbox{\small$p_1;D$},%
               label.side=left,label.dist=10}{i,r2}
  \fmf{fermion,label=\mbox{\small$p_2;A$},%
               label.side=left,label.dist=10}{r1,i}
\fmfblob{40}{i}
\fmfv{label=\mbox{/\hskip 1mm \mbox{\small$\begin{array}{c} 
      C \\[0.7mm] \raisebox{-1mm}{B} \end{array}$} 
      \hskip 1mm $\begin{array}{c} \backslash \\ /
      \end{array}\hskip -4mm$},label.dist=1,label.angle=-165}{i}
\fmffreeze
\fmf{phantom_arrow}{l,i}
\end{fmfgraph*}}
\\*[0.5ex] 
& = \left[ e\gamma^\nu \, \tfrac{\kslash_1 + \mu}{k_1^2-\mu^2} \,
e\gamma^\mu \, \tfrac{\kslash_1 + \mu}{k_1^2-\mu^2} \,
e \gamma_\nu \,\tfrac{1}{k_1^2-M^2} \right]_B^C \times 
\left[ e \gamma^\kappa \, \tfrac{\kslash_2 + \mu}{k_2^2 -\mu^2} 
\right]_A^B\, \left[
\tfrac{\kslash_2 +\mu}{k_2^2 -\mu^2} \, 
e \gamma_\kappa \,  \tfrac{1}{k_2^2-M^2} \right]_C^D~, \notag 
\\[2ex] 
R[ & ((v_1)v_2)] = ~\parbox{30mm}{\begin{fmfgraph*}(30,15)
\fmfleft{l}
\fmfright{r1,r2}
  \fmf{photon,label=\mbox{\small$p_1{-}p_2;\mu$},%
               label.side=right,label.dist=14}{l,i}
  \fmf{fermion,label=\mbox{\small$p_1;D$},%
               label.side=left,label.dist=10}{i,r2}
  \fmf{fermion,label=\mbox{\small$p_2;A$},%
               label.side=left,label.dist=10}{r1,i}
\fmfblob{40}{i}
\fmfv{decor.shape=circle,decor.filled=full,decor.size=40,
      label=\mbox{/\hskip 5mm $\begin{array}{c} \backslash \\ /
      \end{array}\hskip -4mm$},label.dist=1,label.angle=180}{i}
\fmffreeze
\fmf{phantom_arrow}{l,i}
\end{fmfgraph*}}
\\*[0.5ex]
&= \left[ e \gamma^\kappa \, \tfrac{\kslash_2 + \mu}{k_2^2 -\mu^2} \, 
e\gamma^\nu \, \tfrac{\kslash_1 + \mu}{k_1^2-\mu^2} \,
e\gamma^\mu \, \tfrac{\kslash_1 + \mu}{k_1^2-\mu^2} \,
e \gamma_\nu \,\tfrac{1}{(k_1-k_2)^2-M^2} \, 
\tfrac{\kslash_2 +\mu}{k^2 -\mu^2} \, 
e \gamma_\kappa \, \tfrac{1}{k_2^2-M^2} \right]_A^D  \;.\notag 
\end{align}
\end{subequations}%
\end{fmffile}
Both $R[R[(v_1)](v_2)]$ and $R[((v_1)v_2)]$ define local
counterterms, but both integrals $\int d^4 k_2 d^4 k_1 \;
\{ ((v_1)v_2) - R[((v_1)v_2)]\}$ and $\int d^4 k_2 d^4 k_1
\; \{ ((v_1)v_2) - R[R[(v_1)] (v_2)]\}$ are
\emph{infinite}. To obtain a finite expression one has to include
$R[(v_1)] (v_2)$ in a way given by the forest formula.

We must say a few words how equivalence is defined
quantitatively. Renormalization schemes depend on some regularization
parameter $\epsilon$. Infinities correspond to pole terms in
$\epsilon$. In terms of $\epsilon$, Kreimer gives the following
definition of equivalence:
\begin{equation}
X \sim Y \qquad\mbox{iff} \qquad \lim_{\hbar \to 0, \epsilon \to 0} 
\{X-Y\}=0~.
\label{sim}
\end{equation}
Accordingly, $R$ is a renormalization map iff $R[X] \sim X$ for all
PWs $X$. It is important to understand that $R[X] \sim X$ does not
imply $R[X] Y \sim XY$. The reason is that if $Y$ has pole terms in
$\epsilon$ then in the product $(R[X]-X)Y$ also terms of order
$\epsilon$ in $R[X]-Y$ become essential. It turns out that the full
set of properties of a Hopf algebra can only be guaranteed if
equivalence works for products, in a certain sense. The precise
condition to the the renormalization map $R$ is
\begin{equation}
R\big[\tprod_i  R[X_i] \; \tprod_j Y_j \big] 
= \tprod_i  R[X_i] \; \tprod_j R[Y_j] ~. 
\label{equiv}
\end{equation}
We indicate by $X \approx Y$ that under the condition \eqref{equiv} we
have $X \sim Y$, but that in general equivalence is not guaranteed. 

In the BPHZ scheme there is no regularization parameter $\epsilon$, so
we cannot use the definition \eqref{sim}. Nevertheless, $R$ is defined
for any Feynman graph, and we say that $X \sim Y$ iff $Y=X$ or
$Y=R[X]$. The condition \eqref{equiv} makes sense, and we have $R^2=R$
by construction. We remark that superficially convergent graphs with
subdivergences (if included, see the remark at the end of section
\ref{for}) are annihilated by $R$. This is clear in the BPHZ scheme,
because a Taylor expansion until order $d<0$ makes no sense. In what
follows we work on a general level without specifying the
renormalization scheme and its $R$-operation.

\section{The Hopf algebra}
\label{hopf}

Following the work of Kreimer \cite{k} we will now equip the PWs with
the structure of a (quasi-) Hopf algebra. This goes in four
steps. First, we would like to consider the set $\mathcal{A}$ of all
PWs (which include from now on its $R$-equivalents) as a vector
space. We enlarge formally this set $\mathcal{A}$ by
all rational linear combinations of PWs. This makes $\mathcal{A}$ to a
formal vector space over the field $\mathbf{Q}$ of rational numbers,
$\mathbf{Q}$ just for simplicity.

The second step makes $\mathcal{A}$ to an algebra by defining a
product $m$. This is an operation which assigns to a sum of pairs of
elements of $\mathcal{A}$ a new one. Actually only
$\mathbf{Q}$-equivalence classes of pairs are essential so that 
$m$ operates on the tensor product, $m: \mathcal{A} \otimes \mathcal{A}
\to \mathcal{A}$. According to \cite{k} we build
the commutative and associative formal product 
\[
m[X \otimes Y] = XY=YX~,\qquad X,Y \in \mathcal{A}~,
\]
corresponding to two disjoint divergences. We further define a formal
unit $e$ by
\[
m[e \otimes X]=m[X \otimes e] =X \qquad \forall\; X \in \mathcal{A}~.
\]
The unit $e$ is \emph{not} considered as a PSW. It is convenient to
consider $e$ as produced by an operation
\[
E: \mathbf{Q} \to \mathcal{A}~, \qquad E(q) = qe~.
\]

The third step is to make $\mathcal{A}$ to a coalgebra. The operations
of a coalgebra are the duals of the algebra operations. Dual means
turning the arrows. For instance, the dual of the above unit $E$, the
counit $\varepsilon$, will be a formal operation given by
\[
\varepsilon: \mathcal{A} \to \mathbf{Q}~, \qquad \varepsilon[q e] := q~,
\qquad \varepsilon[X] := 0 ~~\forall\, X \neq e~, ~~ X \in  \mathcal{A}~.
\]
Now comes a physically significant ingredient of our coalgebra, the
coproduct $\Delta$. A product was the assignment of one element to
sums of pairs of other elements. Hence, a coproduct will be the
splitting of one element into sums of pairs of other elements, in symbols
\[
\Delta: \mathcal{A} \to \mathcal{A} \otimes \mathcal{A}~.
\]
The philosophy is that $\Delta$ provides the splitting of a 1PI-graph
$\Gamma$ into a formal sum of tensor products of all possible
divergent subgraphs $\gamma_i$ (left factor) by the fraction
$\Gamma/\gamma_i$ obtained by reducing $\gamma_i$ to a hole (right
factor). The left factors are, moreover, treated by the $R$-operation.

Let us formalize this idea. The graph $\Gamma$ is represented by a PW
$X$ describing its forest structure. Let $\{X_i\}$ be a subset of PSWs
of $X$ in the sense of section~\ref{for}. We are going to define the
fraction $X/(\prod_i X_i)$. If $\prod_i X_i =X$ we define
$X/X=e$. Otherwise we label the rows of $X$. Each row of $X_i$ is a
substring of one determined row of $X$. We give to the $X_i$-rows the
labels of the $X$-rows they are contained in. These labels could be
ambiguous but we fix one choice for all subwords of $X$. We delete
from $X$ and all $X_i$ all but those rows whose labels occur in each
of the chosen PSWs $X_i$. Let the results be $X'$ and $X_i'$. If there
remains no row at all or if $X_i' \cap X_j' \neq \emptyset$ for some
pair $\{X_i',X_j'\}$ then we put $X/(\prod_i X_i)=\emptyset.$
Otherwise $X/(\prod X_i)$ is given by removing all $X_i'$ from $X'$.

Now, the coproduct of a PW $X$ containing the PSWs $X_1,\dots X_n$ is
defined by
\begin{align}
\Delta[e] &:= e \otimes e\,, \notag \\
\Delta [X] &:= e \otimes X + \tsum_T \Big\{
\tprod_{i \in T} R[X_i'] \otimes X/(\tprod_{i \in T} X_i) \Big\} \,, 
\end{align}
where the sum runs over all ordered subsets $T=\{i_1,\dots,i_k\}
\subset \{1,2,\dots,n\}$, $i_1<i_2<\dots <i_k$. The order of the
factors and products is not important in this definition, but we must
avoid taking identical terms several times. In the sequel we will omit
the primes on $X_i'$ which indicate the truncation to the common rows.

Our algebra $\mathcal{A}$ also contains elements of the type $R[X]$,
where $X$ is a PW. Kreimer gives two possible definitions for $\Delta
\circ R$,
\begin{subequations}
\begin{align}
\Delta[R[X]] & =\Delta[X] ~, \label{dr1} \\
\Delta[R[X]] &=(\id \otimes R)' \circ \Delta[X]~, \label{dr2}
\end{align}
\end{subequations}
where the prime means that $R[e]$ is replaced by $e$. Kreimer chooses
to work with \eqref{dr1}. This choice violates coassociativity, but
non-coassociativity is interesting from a number theoretical point of
view \cite{kh}. We prefer \eqref{dr2}, because $R[X]$ is always a local
counterterm $\bullet\,$. The philosophy is that $\Delta$ splits a
graph into subgraphs and remainders. Hence, both of them should be
local counterterms in this example, $\Delta[\bullet]=\sum \bullet
\otimes \bullet$, and for us the natural definition is \eqref{dr2} or
\begin{align}
\Delta [R[X]] &:= e \otimes R[X] 
+ \tsum_T \Big\{ \tprod_{i \in T} R[X_i] \otimes 
R[X/(\tprod_{i \in T} X_i)]' \Big\} \,. 
\raisetag{1.5ex}
\end{align}
Again, the prime means that $R[X/X]$ has to be replaced by $e$ instead
of $R[e]$. This can be easily interpreted in terms of PSWs. The PSWs
$X_i$ of $R[X]$ are identical with the PSWs of $X$, except for the
total PW $R[X]$.  The fraction $R[X]/(\prod_i X_i)$ obtained by
removing the PSWs $X_i$ in $R[X]$ clearly coincides with $R[X/\prod_i
X_i]$, except for $R[X]/R[X]=e$.

There are of course some consistency conditions to fulfill before we
can call $\mathcal{A}$ a coalgebra. One of these conditions to
$\Delta$ is coassociativity, which is derived from associativity by
turning the arrows: If we split one element into a sum of pairs, it must
be the same to split the left or the right factor further. In symbols,
coassociativity means
\begin{equation}
(\id \otimes \Delta) \circ \Delta[X] = (\Delta \otimes \id) \circ
\Delta[X] ~,\qquad \forall\; X \in \mathcal{A}~.
\end{equation}
We give the proof in proposition \ref{coass} in the appendix. For the
choice \eqref{dr1}, coassociativity was only satisfied under the
additional condition \eqref{equiv}, but also with \eqref{dr2} we
need \eqref{equiv} to get a true Hopf algebra, see below. 

The `counit' $\varepsilon$ is only a left counit and becomes a
true counit under the condition \eqref{equiv}. Recall that 
an element of $\mathcal{A}$ is a formal linear combination of
products $X=\tprod_i X_i \;\tprod_j R[Y_j]$, where $X_i,Y_j$ are
iPWs. We have 
\[
\Delta [X] = \tprod_i R[X_i] \; \tprod_j R[Y_j] \otimes e 
+ e \otimes \tprod_i X_i \; \tprod_j R[Y_j]  + \tsum Z \otimes Z'~,
\qquad X \neq e~,
\]
where $Z,Z'$ stand for terms which do not contain the unit $e$ and
which are annihilated by $\varepsilon$. Hence, the counit axioms read 
\begin{subequations}
\begin{align}
(\varepsilon \otimes \id) \circ \Delta [X] &= 
\tprod_i X_i \; \tprod_j R[Y_j] = X ~,
\\
(\id \otimes \varepsilon) \circ \Delta [X] &= 
\tprod_i R[X_i] \; \tprod_j R[Y_j] \approx R[X] \sim X~. 
\end{align}
\end{subequations}
In the last line we need \eqref{equiv} to obtain equivalence with
$X$. Moreover, the `antipode' $S$ defined below turns out to require
\eqref{equiv} to be a true antipode.

So far we have equipped $\mathcal{A}$ with the structures of an
algebra and a coalgebra. Both merge to a bialgebra if $\Delta$ is an
algebra homomorphism,
\begin{equation}
\Delta \circ m[X \otimes Y] = (m \otimes m) \circ (\id \otimes \tau
\otimes \id) [\Delta[X] \otimes \Delta[Y]]~, 
\quad \forall\; X,Y \in \mathcal{A}~. 
\label{DXY}
\end{equation}
Here, $\tau[X \otimes Y]:=Y \otimes X$ denotes the flip operation.
It is evident that \eqref{DXY} is fulfilled, because the subwords of
$XY$ are the subwords $X_i$ of $X$ and $Y_i$ of $Y$
together. 

The last step extends the bialgebra to a Hopf algebra. On a Hopf
algebra there exists the additional structure of an antipode
$S:\mathcal{A} \to \mathcal{A}$, which is the dual of the inverse in
an algebra. Our algebra does not have an inverse (except for
$e^{-1}=e$), nevertheless it has (under the condition \eqref{equiv})
an antipode, which will provide the link to the forest formula:
\begin{subequations}
\label{ap}
\begin{align}
S[e] &= e ~,  \\
S[XY] &= S[Y] S[X] ~,\qquad && \forall X,Y \in \mathcal{A}~, 
\label{ap2}  \\
S[X] &= -X - m \circ (\id \otimes S) \circ P_2 \circ \Delta[X] ~,\qquad{} 
&& \forall\mbox{ iPW $X \in \mathcal{A}$}~, \label{ap3} \\
S[R[X]] &= -R[ X + m \circ (S \otimes \id) \circ P_2 \circ \Delta[X]] ~, 
\qquad{} && \forall\mbox{ iPW $X \in \mathcal{A}$}~, \label{ap4}
\end{align}
\end{subequations}
where $P_2 = (\id-E \circ \varepsilon) \otimes (\id-E \circ
\varepsilon)$. The antipode is by \eqref{ap} recursively defined,
because in $P_2 \circ \Delta[X]$ only smaller words than $X$ survive,
and for primitive words $(x)$ we simply have $S[(x)]=-(x)$ and
$S[R[(x)]]=-R[(x)]$. We show in proposition~\ref{anti} that of the
four axioms on $S$ to check, only one is fulfilled in general
renormalization schemes, the other three require \eqref{equiv}:
\begin{subequations}
\begin{align}
m \circ (S \otimes \id) \circ \Delta [X] & \sim E \circ \varepsilon [X] ~,
\label{msi} 
\\
m \circ (\id \otimes S) \circ \Delta [X] & 
\approx E \circ \varepsilon [X]~,
\\
m \circ (S \otimes \id) \circ \Delta [R[X]] & 
\approx E \circ \varepsilon [R[X]] 
\approx 
m \circ (\id \otimes S) \circ \Delta [R[X]] ~.
\end{align}
\end{subequations}
Formula \eqref{msi} relies deeply on the fact that for $X$ being an
iPW, the equation  
\begin{subequations}
\label{bogo}
\begin{align}
m \circ (S \otimes \id) \circ \Delta [X] &= (\id - R) \Big[ 
X + \tsum_T \Big\{ m \Big[\tprod_{i \in T} (-R[\bar{X}_i]) 
\otimes X/(\tprod_{i \in T} \! X_i) \Big] \Big\} \Big] 
\notag \\[-1ex]
&= (\id-R) [\bar{X}]~, \label{bogoa} \\
R[\bar{X}_i] &:= - S[R[X_i]]~, \notag
\end{align}
reproduces Bogoliubov's recurrence formula of renormalization
\cite{bs}. Here, $X_i \neq X$, $i=1,\dots,n$, are the proper PSWs of
$X$. Denoting by $X_{ij} \neq X_i$, $j=1,\dots n_i$, the proper PSW of
$X_i$, we can write
\begin{align}
R[\bar{X}_i] \equiv -S[R[X_i]] &
= R[X_i + m \circ (S \otimes \id) \circ P_2 \circ \Delta[X_i]] \notag
\\ 
&= R \Big[ X_i + \tsum_{T_i} \Big\{ m \Big[ \tprod_{j \in T_i} \!\!
S[R[X_{ij}]] \otimes X_i / (\tprod_{j \in T_i}\!\! X_{ij}) 
\Big] \Big\} \Big]~.
\end{align}
\end{subequations}
Thus, $\bar{X}_i$ has the same structure as $\bar{X}$, and we obtain
indeed a recurrence formula. The integrand $\bar{X}$ associated to an
integrand $X$ is pre-finite, which means that all subdivergences are
compensated. The remaining superficial divergence is compensated by
$\id{-}R$.

To identify \eqref{bogo} with Bogoliubov's recurrence formula it is
important that the coproduct produces all combinations of disjoint
subdivergences, which are encoded in the set of maximal forests.
This means that in describing a Feynman graph $\Gamma$ with
subdivergences by a parenthesized word $X$, we must somehow include in
$X$ all maximal forests of $\Gamma$. That is why we have written the
maximal forests as lines of $X$. The maximal forests are defined by
the relative position of the subdivergences. Each time we meet an
overlap of subdivergences we have a branching of forests. Having
defined the forests we must say how to detect the disjoint
subdivergences. Forests contain by definition no overlapping
divergences, so the only problem is to avoid nested divergences. This
was achieved by our factorization procedure $X/(\prod_{i \in T} X_i)$,
which yields zero if the $X_i$ intersect. By variation of $T$ (which
must be an ordered set to avoid the multiplicities) we get all
products of disjoint subdivergences. It is important that if a
subdivergence occurs in two or more forests, we must count it only
once. That is why we have introduced the brackets connecting identical
regions in various maximal forests. 

In conclusion, our modified definition of a parenthesized word that
keeps track of different maximal forests and connects simultaneously
shrinkable boxes is the correct language for Bogoliubov's recurrence
formula \cite{bs}. This formula has an explicit solution, Zimmermann's
forest formula \cite{z}. Both are reproduced by coproduct and antipode
of a (quasi-) Hopf algebra via $m \circ (S \otimes \id) \circ \Delta$.
We remark that the crucial formula \eqref{msi} is actually a stronger
equivalence $\simeq$. Due to the forest formula \eqref{bogo}, the
difference between left and right hand sides is \emph{finite} in any
renormalization scheme. 

\section{The primitivator $\mathcal{P}$ and the relation to the Hopf
algebra of Kreimer}
\label{KWK}

Having worked out a Hopf algebra of Feynman graphs where overlapping
divergences are treated on the same footing as disjoint and nested
ones, we must also say what the precise relation is to Kreimer's
formulation \cite{k} where overlapping divergences are resolved before
building the Hopf algebra. Our presentation is inspired by an idea of
Dirk Kreimer. A detailed discussion of overlapping divergences based
on set-theoretical reasoning was given in \cite{k3}, some remarks can
also be found in the appendix of \cite{ck}.

The connection to Kreimer's Hopf algebra is achieved by introduction
of a ``primitivator'' $\mathcal{P}$ which maps overlapping divergences
to primitive elements. Let $X$ be an iPW with proper PSWs $X_i \neq
X$, $i=1,\dots,n$, and $T \subset \{1,\dots,n\}$. Let us write the
exterior parentheses of iPWs explicitly, i.e.\ $(X)$ instead of $X$
and $(X_i)$ instead of $X_i$ and $(\mathcal{P}[X/\tprod_{i \in T}
X_i])$ instead of $\mathcal{P}[X/\tprod_{i \in T} X_i]$. With this
convention we define
\begin{equation}
\mathcal{P}[(X)] := (X) - \tsum_{T} \Big( \tprod_{i \in T} (X_i) \:
\mathcal{P}\big[ X/\tprod_{i \in T} X_i\big] \Big)~.
\label{Prim}
\end{equation}
We are going to prove that $\mathcal{P}[(X)]$ is primitive in the
following sense:
\begin{equation}
\Delta[\mathcal{P}[(X)]]= e \otimes \mathcal{P}[(X)] 
+ R[\mathcal{P}[(X)]] \otimes e ~.
\label{DP}
\end{equation}

If $(X)$ is primitive it contains no PSWs. Hence we have $T =
\emptyset$ and $\mathcal{P}[(X)]=(X)$. For $(X)$ and $(Y)$ being
primitive we compute $\mathcal{P}[((Y)X)] = ((Y)X)-((Y)X)=0$. By
induction it is easy to show that $\mathcal{P}[Y]=0$ for any
non-primitive one-line iPW $Y$. To prove \eqref{DP} by
induction we assume that all $(\mathcal{P}\big[ X/\tprod_{i \in T}
X_i\big])$ are primitive in the sense \eqref{DP}. Hence the only PSWs
of $\Big( \tprod_{i \in T} (X_i) \: \mathcal{P}\big[ X/\tprod_{i \in
T} X_i\big] \Big)$ are the $(X_i)$ and their subwords $(X_{k_i})$, with
$k_i \in T^i \subset \{1,\dots n_i\}$. We compute
\begin{align}
\Delta[\mathcal{P}[(X)]] &= e \otimes (X) + R[(X)] \otimes e + 
\tsum_T \tprod_{i \in T} R[(X_i)] \otimes (X/\tprod_{i\in T} X_i) 
\notag \\
& - \tsum_{T} \Big\{ e \otimes \Big(\tprod_{i \in T} (X_i) \:
\mathcal{P}\big[ X/ \! \tprod_{i \in T} X_i\big] \Big) 
+ R\Big[\Big(\tprod_{i \in T} (X_i) \:
\mathcal{P}\big[ X/ \! \tprod_{i \in T} X_i\big] \Big)\Big] \otimes e 
\notag \\
& + \tprod_{i \in T} R[(X_i)] \otimes 
\big( \mathcal{P}\big[ X/\tprod_{i \in T} X_i\big] \big) \Big\}
\label{DPX}
\\
&- \tsum_{T_1,T_2,T_3,\bigcup_{m \in T_2} T^m} \bigg\{ 
\tprod_{i \in T_3} \! R[(X_i)] \tprod_{m \in T_2} \!\! \Big\{ 
\tprod_{k_m \in T^m} \!\!\! R[(X_{k_m})] \Big\} \otimes 
\notag \\[-1ex] & \hskip 7em \notag
\otimes \Big( \tprod_{l \in T_1} \! (X_l) \tprod_{m \in T_2} \!\!
(X_m/ \!\!\! \tprod_{k_m \in T^m} \!\!\! X_{k_m}) 
\;\mathcal{P} \big[ X/ \!\!\!\! \tprod_{j \in T_1\oplus T_2 \oplus
T_3} \!\!\!\! X_j\big] \Big) \bigg\} .
\end{align}
In the last (splitted) line we have $T_1 \oplus T_2 \neq \emptyset$
because that contribution has been written explicitly in the line
before. Comparison of \eqref{xxcc} with \eqref{xxc} in the appendix shows
that the last (splitted) line of \eqref{DPX} equals
\[
-\tsum_{T,T'} \Big\{ \tprod_{i \in T} \! R[(X_i)] \otimes 
\Big( \tprod_{j \in T'} \big( \big\{ X/ \tprod_{i \in T} \! X_i
\big\}_j \big) ~ \mathcal{P}\big[ \big\{ X/ \tprod_{i \in T} \! X_i
\big\} / \big\{ X/ \tprod_{i \in T} \! X_i \big\}_j \big] \Big) \Big\}~,
\]
where $\big\{ X/ \tprod_{i \in T} \! X_i \big\}_j\;$, $j \in T'$, are
the PSWs of $(X/ \tprod_{i \in T} \! X_i)$. Using the definition
\eqref{Prim} for $(X)$ and $(X/ \tprod_{i \in T} \! X_i)$ we confirm
\eqref{DP}.

This means that we may replace the overlapping divergence $(X)$ by the
linear combination $\mathcal{P}[(X)] + \tsum_{T} \Big( \tprod_{i \in
T} (X_i) \: \mathcal{P}\big[ X/\tprod_{i \in T} X_i\big] \Big)$.  If
$(X)$ is an overlapping divergence which contains no overlapping
subdivergences, all $X_i$ are one-line PWs (or connected identical
rows of one-line PWs $\tilde{X}_i\,$; in that case we replace $X_i$ by
$\tilde{X_i}$) . Since the $\mathcal{P}[(X)]$ form additional
primitive (i.e.\ one-line) elements of the Hopf algebra, we have
written the multi-line overlapping divergence $(X)$ as a linear
combination of one-line PWs. In other words, our Hopf algebra is
isomorphic to a Hopf algebra of one-line PWs, and this is precisely
Kreimer's original Hopf algebra. The primitive elements of Kreimer's
Hopf algebra are the graphically primitive elements and from each
overlapping divergence a computational-primitive element. Our approach
provides an explicit construction of the latter. The same can be
achieved, for instance, by Schwinger-Dyson techniques \cite{k,kh} or
set-theoretical considerations \cite{k3}.

The advantage of Kreimer's Hopf algebra of one-line PWs is that it can
be reformulated as a Hopf algebra of rooted trees \cite{ck}. A
subalgebra thereof turns out to be the dual of the diffeomorphism
group of a manifold. It is now interesting to ask \cite{ck} for the 
(noncommutative) manifold whose diffeomorphism group is the dual of
the Hopf algebra of renormalization. We feel that answering this 
question is indispensable for a true understanding of renormalization
and of the short-distance structure of spacetime.

\section{Two examples for the coproduct and the forest formula}
\label{ex2}

We compute here the coproducts and forest formulas for two striking 
examples of section \ref{ex1}. By PSW we shall always mean proper PSW,
we write the trivial PWs explicitly. The proper PSWs of
\begin{gather}
X= \begin{array}{rl} 
((v_1) p_2) &\VL \\ ((v_2) p_1) &\AL \end{array} 
\qquad \qquad \qquad \parbox{30mm}{\fmfreuse{polar}}
\tag{\ref{PV}}
\\
\parbox{\textwidth}{are obviously} \notag
\\
X_1=(v_1)\,,\qquad X_2 = (v_2)\,. \tag{$\ref{PV}_\mathrm{s}$}
\end{gather}
Let us compute $X/X_1$. The only row of $X_1$ can only be related to
the upper row of $X$ so that $X'=((v_1)p_2)$. To obtain $X/X_1$ we
must remove $X_1$ from $X'$, the result is $X/X_1=(p_2)$. Accordingly,
\begin{equation}
X/X_1=(p_2)\,,\qquad X/X_2=(p_1)\,,\qquad X/(X_1 X_2)=0\,.
\tag{$\ref{PV}_\mathrm{r}$}
\end{equation}
The last equation holds because $X_1,X_2$ have no common row
label. Therefore, the coproduct reads
\begin{align}
\Delta [X]
= e \otimes \begin{array}{rl} ((v_1) p_2) &\VL \\ ((v_2) p_1) &\AL 
\end{array}
+ R\! \left[\begin{array}{rl} ((v_1) p_2) &\VL \\ ((v_2) p_1) &\AL 
\end{array} \right] \!\! \otimes e  
 + R[(v_1)] \otimes (p_2) + R[(v_2)] \otimes (p_1) \,.
\tag{$\ref{PV}_\Delta$}
\end{align}
Let us now apply the operator $m \circ (S \otimes \id)$. To avoid
unnecessary calculation we use the general result \eqref{bogoa}, 
\[
m \circ (S \otimes \id) \circ \Delta [X] = 
(\id-R)[X + m \circ (S \otimes \id) \circ P_2 \circ \Delta[X]]~.
\]
The projection $P_2$ removes all terms containing the unit $e$ so that 
in our case we have $P_2 \circ \Delta[X]] =  R[(v_1)] 
\otimes (p_2) + R[(v_2)] \otimes (p_1)$. This gives 
\begin{align}
m \circ {} & {} (S \otimes \id) \circ \Delta [X] \notag \\
& = (\id -R) \left[ \begin{array}{rl} ((v_1) p_2) &\VL \\ 
((v_2) p_1) &\AL \end{array}
+ S[R[(v_1)]] (p_2) 
+ S[R[(v_2)]] (p_1) \right]
\notag 
\\
&= (\id-R)\left[\begin{array}{rl} ((v_1) p_2) &\VL \\ ((v_2) p_1) &\AL 
\end{array} - R[(v_1)] (p_2) - R[(v_2)] (p_1)\right]~.
\tag{$\ref{PV}_\mathrm{f}$}
\end{align}
The primitivator of $X$ reads
\begin{align}
o_1 :=\mathcal{P}[X] = \begin{array}{rl} ((v_1) p_2) &\VL \\ ((v_2) p_1) &\AL 
\end{array} - ((v_1)p_2) - ((v_2)p_1)~.
\tag{$\ref{PV}_\mathrm{p}$} 
\label{o1}
\end{align}
It is easy to verify $\Delta[o_1]=R[o_1] \otimes e + e \otimes o_1$.

Let us do the same steps for example \eqref{PVV}: 
\begin{align}
X &=\begin{array}{rcl} 
(((v_3) \VL& v_{13}) p_2) &\VL \\ 
(((v_3) \AL& v_{23}) p_1) &\AL \end{array} \,,
\hskip 3cm
\parbox{30mm}{\fmfreuse{fmfPVV}}
\tag{\ref{PVV}}
\\
X_1 &=\begin{array}{r} (v_3) \VL \\ (v_3) \AL \end{array} \,, \qquad  
X_2 = ((v_3)v_{13})\,, \qquad 
X_3= ((v_3)v_{23})\,, 
\tag{$\ref{PVV}_\mathrm{s}$}
\\
X/X_1 &= \begin{array}{rcl} ((v_{13}) p_2) &\VL \\ 
((v_{23}) p_1) &\AL \end{array} \,, \quad
X/X_2=(p_2)\,,\qquad  X/X_3 = (p_1)\,,
\tag{$\ref{PVV}_\mathrm{r}$}
\\*
&X/(X_1X_2)=X/(X_1X_3)=X/(X_2X_3)=X/(X_1X_2X_3)=0\,, \notag 
\\[1ex]
\Delta [X] &= e \otimes \begin{array}{rcl} 
(((v_3) \VL& v_{13}) p_2) &\VL \\ 
(((v_3) \AL& v_{23}) p_1) &\AL \end{array} 
+ R\left[ \begin{array}{rcl} 
(((v_3) \VL& v_{13}) p_2) &\VL \\ 
(((v_3) \AL& v_{23}) p_1) &\AL \end{array} \right] \otimes e 
+ R[(v_3)] \otimes \begin{array}{rr} ((v_{13}) p_2) &\VL \\ 
((v_{23}) p_1) &\AL \end{array} 
\notag \\*
&+ R[((v_3)v_{13})] \otimes (p_2) 
+ R[((v_3)v_{23})] \otimes (p_1)\,,
\tag{$\ref{PVV}_\Delta$}
\\*
& \hskip -10mm \big(\mbox{in the third term, 
$\begin{array}{r} (v_3) \VL \\ 
(v_3) \AL \end{array}$ can be condensed to $(v_3)$}\big) \notag 
\\[3ex]
m \circ {}&{}(S \otimes \id) \otimes \Delta[X] \notag 
\\*
&= (\id-R) \bigg[
\begin{array}{rcl} 
(((v_3) \VL& v_{13}) p_2) &\VL \\ 
(((v_3) \AL& v_{23}) p_1) &\AL \end{array} 
+ S [R[(v_3)]] \, \begin{array}{rl} ((v_{13}) p_2) &\VL \\ 
((v_{23}) p_1) &\AL \end{array} 
\notag \\[-1ex]
&\hskip 10mm + S\big[R\big[ ((v_3)v_{13}) \big] \big] (p_2) 
+ S\big[ R\big[ ((v_3)v_{23})\big] \big] (p_1) \bigg]
\notag 
\\
&= (\id-R) \bigg[
\begin{array}{rcl} 
(((v_3) \VL& v_{13}) p_2) &\VL \\ 
(((v_3) \AL& v_{23}) p_1) &\AL \end{array} 
- R[(v_3)] \, \begin{array}{rl} ((v_{13}) p_2) &\VL \\ 
((v_{23}) p_1) &\AL \end{array} 
\notag \\*
& \hskip 10mm - \Big\{ R[((v_3)v_{13})] 
+ R\big[ m \circ (S \otimes \id) \circ P_2
\Delta [((v_3)v_{13})] \big] \Big\} (p_2) 
\notag \\*
& \hskip 10mm - \Big\{ R[((v_3)v_{23})] 
+ R \big[ m \circ (S \otimes \id) \circ P_2
\Delta [((v_3)v_{23})] \big] \Big\} (p_1) \bigg] \notag 
\\
&= (\id-R) \bigg[
\begin{array}{rcl} 
(((v_3) \VL& v_{13}) p_2) &\VL \\ 
(((v_3) \AL& v_{23}) p_1) &\AL \end{array} 
- R[(v_3)] \, \begin{array}{rl} (v_{13}) p_2) &\VL \\ 
(v_{23}) p_1) &\AL \end{array} 
\notag \\*
&\hskip 10mm - R[((v_3)v_{13})] (p_2) 
+ R \big[ R[(v_3)] (v_{13}) \big] (p_2) 
\notag \\*
&\hskip 10mm - R[((v_3)v_{23})] (p_1) 
+ R \big[ R[(v_3)] (v_{23}) \big] (p_1) \bigg]~,
\tag{$\ref{PVV}_\mathrm{f}$}
\\
o_2 & :=\mathcal{P}[X]=\begin{array}{rcl} 
(((v_3) \VL& v_{13}) p_2) &\VL \\ 
(((v_3) \AL& v_{23}) p_1) &\AL \end{array} 
- ((v_3)o_1) - (((v_3)v_{13}) p_2) - (((v_3)v_{23}) p_1) ~.
\tag{$\ref{PVV}_\mathrm{p}$}
\end{align}
The primitive element $o_1$ computed in \eqref{o1} enters the
decomposition of $X$ into one-line PWs. 

Example \eqref{VVP} is similar to \eqref{PVV} and is left as an
exercise to the reader. Example \eqref{phi4} is the obvious
generalization of \eqref{PV} to three maximal forests.

\section*{Acknowledgments}

We are grateful to Dirk Kreimer for explaining us the way he treats
overlapping divergences and for discovering the link between our Hopf
algebras. We would like to thank Bruno Iochum, Ctirad
Klim\v{c}ik, Serge Lazzarini and Thomas Sch\"ucker for discussions.

\section*{Appendix: Verification of the Hopf algebra properties}
\setcounter{equation}{0}
\renewcommand{\theequation}{A.\arabic{equation}}

\begin{prp} 
The coproduct $\Delta$ is coassociative, 
$(\Delta \otimes \id) \circ
\Delta = (\id \otimes \Delta) \circ \Delta$. 
\label{coass}
\end{prp}
\textit{Proof.}  Let $X$ be an iPW which is not $R[X']$. Let $X_i \neq
X$, $i=1,\dots,n$, be the proper PSW of $X$. Let $T$ be the set of all
ordered subsets of $\{1,2,\dots,n\}$. We write the contribution
of the trivial PSW $X$ of $X$ explicitly:
\[
\Delta[X]=e \otimes X + R[X] \otimes e + \tsum_T \Big\{
\tprod_{i \in T} R[X_i]
\otimes X/(\tprod_{i \in T} X_i) \Big\}~.
\]
This gives 
\begin{align}
&(\id \otimes \Delta) \circ \Delta[X] \label{iddelta} \\*
&= e \otimes \Big\{ e \otimes X + R[X] \otimes e 
+ \tsum_T \Big\{ \tprod_{i \in T} R[X_i] \otimes 
X/(\tprod_{i \in T}X_i) \Big\} \Big\} 
+ R[X] \otimes e \otimes e \notag 
\\*
&+ \tsum_T \Big\{ \tprod_{i \in T} R[X_i] \otimes 
e \otimes X/(\tprod_{i \in T}X_i) \Big\}
+ \tsum_T \Big\{ \tprod_{i \in T} R[X_i] \otimes 
R[X/(\tprod_{i \in T} X_i) ] \otimes e \Big\} \notag 
\\*
& + \! \tsum_T \!\! \Big\{ \! \tprod_{i \in T} \!\! 
R[X_i] \otimes \!
\tsum_{T'} \!\! \Big\{ \! \tprod_{j \in T'} \!\!\! R \big[ 
\big\{X/(\tprod_{i \in T} X_i) \big\}_{\!j} \big] 
\otimes \big\{X/(\tprod_{i \in T} \! X_i) \big\} /
\big( \! \tprod_{j \in T'} \big\{X/(\tprod_{i \in T} \! X_i) 
\big\}_{\!j} \big) \! \Big\} \! \Big\}, \notag
\end{align}
where $\big\{X/(\tprod_{i \in T} X_i) \big\}_{\!j}$ are the proper PSW of
$X/(\tprod_{i \in T} X_i)$, $j=1,\dots,n'<n,$ and $T'$ is the set
of all ordered subsets of $\{1,\dots,n'\}$. 
The following terms can be
rearranged:
\begin{align}
&e \otimes e \otimes X \notag \\[-1ex]
&+ \big\{ e \otimes R[X] \otimes e + R[X] \otimes e \otimes e 
+ \tsum_T \Big\{ \tprod_{i \in T} R[X_i] \otimes 
R[X/(\tprod_{i \in T} X_i)] \Big\} \otimes e \big\} \notag
\\[-1ex] 
&= (\Delta \otimes \id) (e \otimes X + R[X] \otimes e) \label{xxa}
\end{align}
so that there remain 
\begin{align}
&\tsum_T \! \Big\{ \! \tprod_{i \in T} \! R[X_i] \otimes e 
\otimes X/(\tprod_{i \in T} \! X_i) \! \Big\}
+ e \otimes \tsum_T \! \Big\{ \!
\tprod_{i \in T} \! R[X_i] \otimes X/(\tprod_{i \in T} \! X_i) 
\! \Big\}
\quad \mbox{and} \label{xxb}
\\
& \tsum_{T,T'} \!\! \Big\{ \! \tprod_{i \in T} \!\! 
R[X_i] \otimes
\!\! \Big\{ \! \tprod_{j \in T'} \!\!\! R \big[ 
\big\{X/(\tprod_{i \in T} X_i) \big\}_{\!j} \big] 
\otimes \big\{X/(\tprod_{i \in T} \! X_i) \big\} /
\big( \tprod_{j \in T'}  \big\{X/(\tprod_{i \in T} \! X_i) 
\big\}_{\!j} \big) \! \Big\} \! \Big\} .
\label{xxc}
\end{align}
We investigate $\big\{X/(\tprod_{i \in T} X_i) \big\}_{\!j}$. Either this
is a PSW of $X$ or not. If not there must exist a PSW $X_m$ of $X$ and
some PSWs $X_k$ with $k \in T^m \subset T$ such that
$\big\{X/(\tprod_{i \in T} X_i) \big\}_{\!j} =X_m/(\prod_{k \in T^m}
X_k)$. This means that $T'=T_1 \oplus T_2$ (both $T_1,T_2$ can be
empty but not the sum) and 
\[
\tprod_{j \in T'} \! R\big[ \big\{ X/(\tprod_{i \in T} X_i) 
\big\}_{\!j} \big] = 
\tprod_{l \in T_1} \!\!  R[X_l] \tprod_{m \in T_2} \!\! R[X_m/
(\tprod_{k_m \in T^m} \!\!\! X_{k_m})]~.
\]
Let us assume that $T_2$ contains at least two elements $m_1,m_2$ and
perform the factorization
\begin{equation}
\big\{X/(\tprod_{i \in T} \! X_i) \big\} /
\big( \big\{X_{m_1}/(\tprod_{k_1 \in T^{m_1}} \!\!\! X_{k_1}) \big\} 
\big\{ X_{m_2}/(\tprod_{{k_2} \in T^{m_2}} \!\!\! X_{k_2} )\big\} 
\big)~.
\label{fac}
\end{equation}
Recall that $T^{m_1} {\subset} T$ and $T^{m_2} {\subset} T$ and assume
that $X_n \in T^{m_1} \cap T^{m_2}$. The fraction \eqref{fac} will
only be non-zero if $X_{m_1}/(\prod_{k_1 \in T^{m_1}} \! X_{k_1})$ and
$X_{m_2}/(\prod_{k_2 \in T^{m_2}} \! X_{k_2})$ occur together and
disjoint in at least one row of $X/(\tprod_{i \in T} X_i)$. These rows
correspond to those rows of $X$ each of which contain all $X_i$,
$i{\in} T$, too. But each $X_i$ occurs precisely once in any row, so
does the $X_n$ in question, hence it will either occur in $T^{m_1}$ or
in $T^{m_2}$, but never in both. Therefore, we have a direct sum
decomposition $T=T_3 \oplus \bigoplus_{m \in T_2} T^{m}$ and
\eqref{xxc} takes the form
\begin{align}
\eqref{xxc} &= 
\tsum_{\{T_1,T_2,T_3,\bigcup_{m \in T_2} \!\!T^m\}} \!\! \Big\{ 
\tprod_{i \in T_3}\!\!R[X_i] \tprod_{m \in T_2} \!\! 
\{\tprod_{k_m \in T^m} \!\!\!R[X_{k_m}] \;\} \otimes 
\notag \\[-1ex] 
&{}\qquad \!\! \otimes 
\tprod_{l \in T_1} \!\! R [X_l] \tprod_{m \in T_2} \!\!\!
R\big[X_m/(\tprod_{k_m \in T^m} \!\!\! X_{k_m})\big] 
\otimes X/(\tprod_{m \in T_2} \!\! X_m 
\tprod_{l \in T_1} \!\! X_l 
\tprod_{i \in T_3} \!\! X_i ) \Big\} 
\label{xxcc} \\ \notag 
&= \tsum_T \Big\{\big\{ \tsum_{T_3 \subset T} \tprod_{i \in T_3} \!\! 
R[X_i] \otimes \tsum_{T_1 \subset T/T_3} \tprod_{l \in T_1} \!\! 
R[X_l] \big\} \times 
\\ \notag
& {} ~~ \times \big\{\!  \tprod_{m \in T_2 = T/(T_1\oplus T_3)} 
\tsum_{T^m} \!
\Big\{ \! \tprod_{k_m \in T^m} \!\!\!\!\! R[X_{k_m}] \otimes 
R[X_m/( \!\! \tprod_{k_m \in T^m} \!\!\!\!\! X_{k_m}) ] \Big\} 
\big\} \otimes 
X/ (\tprod_{j \in T} \! X_j ) \Big\} . 
\end{align}
Note that $T_1,T_2,T_3$ can be empty, in that case the missing product
over $R[X_j]$ has to be replaced by $e$. If $T_2$ is empty then the
sum over $T_1=T/T_3$ has to be omitted. Observe that neither $T_1
\oplus T_2$ nor $T_3 \oplus T_2$ can be empty, but these two terms
$T_2= \emptyset$ and either $T_1=\emptyset$ or $T_3=\emptyset$ are
precisely those of \eqref{xxb}. All together can be rewritten as
\begin{align}
\eqref{xxb} + \eqref{xxc} =& \tsum_T \Big\{ \tprod_{j \in T} 
\big\{ e \otimes R[X_j] + R[X_j] \otimes e \: + \notag 
\\[-0.8ex] 
&{}\qquad 
+ \tsum_{T^j} \Big\{ \! \tprod_{k_j \in T^j} \!\!\! R[X_{k_j}] 
\otimes R[X_j/(\tprod_{k_j \in T^j} \!\!\! X_{k_j}) ] 
\Big\} \big\} \otimes 
X/ (\tprod_{j \in T} \! X_j ) \Big\} 
\notag
\\
=& (\Delta \otimes \id) \big[ \tsum_T \big\{ \tprod_{j \in T} R[X_j] 
\otimes X/ (\tprod_{j \in T} \!X_j ) \big\}\big]\,, \label{xxd}
\end{align}
and we conclude 
\begin{equation}
\eqref{xxa}+\eqref{xxb} + \eqref{xxc} = (\Delta \otimes \id) \circ
\Delta[X] = (\id \otimes \Delta) \circ \Delta[X]~.
\end{equation} 

To finish the proof of coassociativity of $\Delta$ we must write down 
\begin{align*}
(\id \otimes \Delta) \circ \Delta[R[X]] 
&= (\id \otimes \Delta) \circ (\id \otimes R') \circ \Delta[X] \\
&= (\id \otimes \id \otimes R') \circ (\id \otimes \Delta) \circ
\Delta[X] \\
&= (\id \otimes \id \otimes R') \circ (\Delta \otimes \id) \circ
\Delta[X] \\
& = (\Delta \otimes \id) \circ \Delta[R[X]] ~,
\\[1ex]
(\id \otimes \Delta) \circ \Delta[XY] 
&=  \hat{m} \big[\{(\id \otimes \Delta) \circ \Delta[X]\} \otimes 
\{(\id \otimes \Delta) \circ \Delta[Y] \} \big] \\
&= \hat{m} \big[ \{(\Delta \otimes \id) \circ \Delta[X]\} \otimes 
\{(\Delta \otimes \id) \circ \Delta[Y] \} \big] \\
&= (\Delta \otimes \id) \circ \Delta[XY] ~.
\end{align*}
We have defined $\hat{m} [ \{ X' {\otimes} X'' {\otimes} X'''\}
\otimes \{ Y' {\otimes} Y'' {\otimes} Y'''\}] := 
X'Y' \otimes X'' Y'' \otimes X''' Y'''$ as well as $R'[e]=e$ and 
$R'[X]=R[X]$ for $X \neq e$. \qed

\begin{prp}
The `antipode' $S$ fulfills $m \circ (S \otimes \id) \circ \Delta
\approx E \circ \varepsilon \approx m \circ (\id \otimes S) \circ
\Delta$, and on PWs $X$ not containing $R$ we even have $m \circ (S
\otimes \id) \circ \Delta [X] \sim 0 = E \circ \varepsilon [X]$.
\label{anti}
\end{prp}
\textit{Proof.}
The case $X=e$ is trivial. Let $X \neq e$ be an iPW, which is not $R[X']$: 
\begin{align*}
m \circ {}&{} (S \otimes \id) \circ \Delta[X] 
= m \circ (S \otimes \id) [ e \otimes X + R[X] \otimes e 
+ P_2 \Delta[X]]
\\
&= X + S[R[X]] + m \circ (S \otimes \id) \circ P_2\Delta[X] 
\\
&= X - R\big[ X + m \circ (S \otimes \id) \circ P_2 \Delta[X] \big] 
+ m \circ (S \otimes \id) \circ P_2\Delta[X] 
\\
&= (\id - R) \big[ X + m \circ (S \otimes \id) \circ P_2\Delta[X]
\big] 
\\
& \sim 0 = E \circ \varepsilon[X]~, 
\\[1ex] 
m \circ {}&{} (\id \otimes S) \circ \Delta[X] 
= m \circ (\id \otimes S) [ e \otimes X + R[X] \otimes e 
+ P_2\Delta[X]] 
\\
&= S[X] + R[X] + m \circ (\id \otimes S) \circ P_2\Delta[X] 
\\
&= -(X + m \circ (\id \otimes S) \circ P_2 \Delta[X])
+R[X] + m \circ (\id \otimes S) \circ P_2\Delta[X] 
\\
&= -(\id - R)[X] \sim 0 = E \circ \varepsilon[X]~. 
\end{align*}
As we have chosen $\eqref{dr2}$, we must also compute ($X$ is again an
iPW)
\begin{align*}
m \circ {}&{} (S \otimes \id) \circ \Delta[R[X]] 
= m \circ (S \otimes \id) [ e \otimes R[X] + R[X] \otimes e 
+ P_2\Delta[R[X]]]
\\
&= R[X] + S[R[X]] + m \circ (S \otimes \id) \circ P_2\Delta[R[X]] 
\\
&= R[X] - R\big[ X + m \circ (S \otimes \id) \circ P_2 \Delta[X] \big] 
+ m \circ (S \otimes \id) \circ P_2\Delta[R[X]] 
\\
& = \big( m \circ (\id \otimes R)- R \circ m \big) \big[
(S \otimes \id) \circ P_2 \Delta[X] \big] 
\\
& \approx 0 = E \circ \varepsilon[R[X]]~.
\end{align*}
We need condition \eqref{equiv} in the form $R \circ m = R \circ m
\circ (\id \otimes R)$ to have equivalence. The remaining case is more
complicated:
\begin{align}
m \circ {}&{} (\id \otimes S) \circ \Delta[R[X]] 
= S[R[X]] + R[X] + m \circ (\id \otimes S) \circ P_2\Delta[R[X]] 
\notag \\
&=- R\big[ X + m \circ (S \otimes \id) \circ P_2 \Delta[X] \big] 
+ R[X] + m \circ (\id \otimes S) \circ P_2\Delta[R[X]] 
\notag \\
&= m \circ (\id \otimes S) \circ P_2\Delta[R[X]] 
- R\big[ m \circ (S \otimes \id) \circ P_2\Delta[X] \big] ~.  \label{ss1}
\end{align}
We transform the first term, using the definition of
$S$ acting on $R[\,.\,]$:
\begin{align}
m {}&{} \circ (\id \otimes S) \circ P_2 \Delta[R[X]]  \label{ss2} \\
&= - m [ P_2 \Delta[R[X]]] 
- m \circ (\id \otimes \{R \circ m \circ (S \otimes \id) \circ P_2
\Delta \}) \circ P_2 \Delta[X] \notag  
\\
&= \big( R \circ m - m \circ (\id \otimes R)\big) \!\circ\! 
\big( P_2 \Delta[X] + (\id \otimes \{m \circ (S \otimes \id) \circ P_2
\Delta \}) \circ P_2 \Delta[X] \big) \notag  
\\
& -R \Big[ m[P_2 \Delta[X]]  
+ m \circ (\id \otimes m) \circ (\id \otimes S \otimes \id) 
\circ (\id \otimes P_2 \Delta) \circ P_2 \Delta[X] \Big] ~.\notag
\end{align}
Now observe that due to coassociativity of $\Delta$ we have 
\begin{align*}
(\id \otimes P_2 \Delta) \circ P_2 \Delta[X] 
&= P_3 \circ (\id \otimes \Delta) \circ \Delta[X] 
= P_3 \circ (\Delta \otimes \id) \circ \Delta[X] \\
&= (P_2 \otimes \id) \circ(\Delta \otimes \id) \circ P_2 \Delta[X] ~,
\end{align*}
with $P_3=(\id {-} E \circ \varepsilon) \otimes (\id {-} E \circ
\varepsilon) \otimes (\id {-} E \circ \varepsilon)$.  Note that
$\Delta$ is multiplicative, not $(P_2\Delta)$. Using also
associativity of $m$ we can write
\begin{align*}
- R \big[ m \circ {}&{} (\id \otimes m) \circ (\id \otimes S 
\otimes \id) 
\circ (\id \otimes P_2 \Delta) \circ P_2 \Delta[X] \big]
\\
&= -R\big[ m \circ (m \otimes \id) \circ (\id \otimes S \otimes \id) 
\circ (P_2 \otimes \id) \circ (\Delta \otimes \id) \circ 
P_2 \Delta[X] \big]~.
\end{align*}
We have computed $(\Delta \otimes \id) \circ P_2 \Delta[X]$ in
\eqref{xxd}. By inspection of that formula we find that $(P_2 \otimes
\id) \circ (\Delta \otimes \id) \circ P_2 \Delta[X]$ equals $(\Delta
\otimes \id) \circ P_2 \Delta[X] -\eqref{xxb}$, which gives
\begin{align*}
- R \big[ & m \circ (\id \otimes m) \circ (\id \otimes S 
\otimes \id) \circ (\id \otimes P_2 \Delta) \circ P_2 \Delta[X] \big] 
\\
&= -R\Big[ \tsum_T m \Big[ \tprod_{j \in T} 
\big\{m \circ (\id \otimes S) \circ \Delta [R[X_j]] \big\} 
\otimes X/\tprod_{j \in T} \! X_j \Big] \Big]
\\
&+ R\Big[ \tsum_T m \Big[ \!\tprod_{j \in T} \! S[[R[X_j]]] 
\otimes X/ \! \tprod_{j \in T} \! X_j \Big] \Big]
+ R\Big[ \tsum_T m \Big[ \! \tprod_{j \in T} \! [R[X_j]] 
\otimes X/ \! \tprod_{j \in T} \! X_j \Big] \Big] \,.
\end{align*}
The last term cancels $-R[m[P_2 \Delta[X]]]$ in \eqref{ss2}
and the middle term cancels $- R[m \circ (S \otimes \id) \circ
P_2\Delta[X]]$ in \eqref{ss1}. We end up with the same problem as
before, to calculate $m \circ (\id \otimes S) \circ \Delta[R[X_i]]$,
however, these $X_i$ are \emph{smaller} than the original $X$. This
leads to an iteration which stops if $X_i$ is primitive, and for primitive
$X_i$ we have 
\[
m \circ (\id \otimes S) \circ \Delta[R[X_i]] = S[R[X_i]]+R[X_i]=0~.
\]
The conclusion is that it is condition \eqref{equiv} required in
\eqref{ss2} which separates us from zero: 
$m \circ (\id \otimes S) \circ \Delta[R[X]] \approx 0 = 
E \circ \varepsilon [R[X]]$ for all iPW $X$.

It remains to apply $m \circ (\id \otimes S) \circ \Delta$ and $m
\circ (S \otimes \id) \circ \Delta$ to products $X=\tprod_i X_i
\tprod_j R[Y_j]$. Here we have the multiplicativity of $\Delta$
\eqref{DXY} and $S$ \eqref{ap2} at disposal, so we clearly get  
\begin{align}
m \circ (\id \otimes S) \circ \Delta [X] & \approx 0=E \circ
\varepsilon [X] \approx  
m \circ (S \otimes \id) \circ \Delta [X] ~.
\end{align}
One case however is special. For $X=\prod X_i$, where none of the $X_i$ is 
$R[X_i']$, we have 
\begin{align}
m \circ (S \otimes \id) \circ \Delta [\tprod_i X_i] &= 
\tprod_i \big\{ (\id - R) \big[ X_i + m \circ (S \otimes \id) 
\circ P_2\Delta[X_i]\big] \big\} \notag \\ 
& \sim 0 = E \circ \varepsilon [\tprod_i X_i]~. \label{S2}
\end{align}
The reason is that $(\id - R) \big[ X_i + m \circ (S \otimes \id)
\circ P_2\Delta[X_i]\big]$ is \emph{convergent as it reproduces the
forest formula}, see \eqref{bogo}. Now, multiplication of $(\id - R)
\big[ X_i + m \circ (S \otimes \id) \circ P_2\Delta[X_i]\big]$ by a
convergent term is equivalent to zero. It is even strongly equivalent
$(\simeq)$ to zero which means that the integral is \emph{finite}. On
the other hand, $m \circ (\id \otimes S) \circ \Delta [\tprod_i X_i] =
\tprod_i \big\{ (R-\id) [ X_i] \big\}$ is a product of divergent
terms, so we need \eqref{equiv} in this case to obtain equivalence to
zero. The fact that $m \circ (S \otimes \id) \circ \Delta$ gives the
forest formula is essential for \eqref{S2} holding in any
renormalization scheme. \qed

\end{document}